\documentclass[10pt,journal]{IEEEtran}
\usepackage{amsmath,amsfonts}
\usepackage{algorithm}
\usepackage{algpseudocode}
\usepackage{array}
\usepackage{tikz}
\usetikzlibrary{positioning}
\usepackage[caption=false,font=footnotesize,labelfont=sf,textfont=sf]{subfig}
\usepackage[colorlinks=true, linkcolor=red, citecolor=red, urlcolor=blue]{hyperref}
\usepackage{textcomp}
\usepackage{stfloats}
\usepackage{diagbox}
\usepackage{url}
\usepackage{verbatim}
\usepackage{graphicx}
\usepackage{cite}
\usepackage{bm}
\usepackage{booktabs}
\usepackage{tabularx}
\usepackage{tikz}
\usetikzlibrary{positioning,arrows.meta,fit,calc,shapes.geometric}
\usepackage{amsthm}
\usepackage{proof}
\usepackage{amssymb}
\usepackage{color}
\usepackage{dsfont} 

\usepackage[most]{tcolorbox}
\definecolor{waraNavy}{HTML}{08236C}
\definecolor{waraBlue}{HTML}{0D5BCB}
\definecolor{waraLane}{HTML}{F2F7FF}
\definecolor{waraSideTop}{HTML}{225BAD}
\definecolor{waraSideBottom}{HTML}{153F86}
\definecolor{waraGreen}{HTML}{EAF7EF}
\definecolor{waraGreenLine}{HTML}{16803A}
\definecolor{waraTag}{HTML}{EEF5FF}
\definecolor{waraTagLine}{HTML}{7DB1FF}
\definecolor{waraOrange}{HTML}{9B4D00}
\definecolor{waraRepair}{HTML}{C0392B}
\usepackage{etoolbox}
\makeatletter
\patchcmd{\@makecaption}
{{\normalfont\footnotesize\scshape #2}}
{{\normalfont\footnotesize #2}}
{}{}
\makeatother
\newtcolorbox{warabox}[1]{
	breakable,
	colback=gray!3,
	colframe=black!55,
	boxrule=0.45pt,
	arc=1pt,
	left=5pt,
	right=5pt,
	top=5pt,
	bottom=5pt,
	fonttitle=\bfseries\footnotesize,
	fontupper=\scriptsize,
	coltitle=black,
	colbacktitle=gray!18,
	title={#1},
	before skip=6pt,
	after skip=6pt
}

\definecolor{litPurpleLine}{HTML}{7B5DD6}
\definecolor{litRedLine}{HTML}{C0392B}

\allowdisplaybreaks
\def\BibTeX{{\rm B\kern-.05em{\sc i\kern-.025em b}\kern-.08em
		T\kern-.1667em\lower.7ex\hbox{E}\kern-.125emX}}
\begin{document}
	
	\title{WARA: Toward Automated Wireless Optimization Research with Closed-Loop LLM Agents}
	
	\author{Yuan Guo,~\IEEEmembership{Member,~IEEE,}
		Yilong~Chen,~\IEEEmembership{Member,~IEEE,}
		Chao~Hu,~\IEEEmembership{Graduate Student Member,~IEEE,}
		Xianghao~Yu,~\IEEEmembership{Senior Member,~IEEE,}
		Liang~Hong,
		and~Jie~Xu,~\IEEEmembership{Fellow,~IEEE}
		\IEEEcompsocitemizethanks{\IEEEcompsocthanksitem Yuan Guo, Yilong Chen, Chao~Hu, and Jie Xu are with the School of Science and Engineering (SSE), the Shenzhen Future Network of Intelligence Institute (FNii-Shenzhen), and the Guangdong Provincial Key Laboratory of Future Networks of Intelligence, The Chinese University of Hong Kong, Shenzhen, Guangdong 518172, China (e-mail: guoyuan@cuhk.edu.cn, chenyilong@cuhk.edu.cn, chaohu@link.cuhk.edu.cn, xujie@cuhk.edu.cn).
			
			Xianghao Yu is with the Department of Electrical Engineering, City University of Hong Kong, Hong Kong (e-mail: alex.yu@cityu.edu.hk).
			
			Liang Hong is with Sun Yat-Sen University, Guangzhou, Guangdong 510006, China (e-mail: hongliang@sysu.edu.cn).
			
			(\textit{Corresponding author: Jie Xu}).}
	}

	\maketitle


	\begin{abstract}
		Large language model (LLM) agents are increasingly capable of external tool use, code generation and execution, intermediate artifact inspection, and iterative output revision, creating new opportunities for automating scientific and engineering research. To the best of our knowledge, this paper presents the first end-to-end autoresearch framework for the wireless domain, with a particular focus on wireless resource allocation optimization, an essential area for characterizing the fundamental performance limits of wireless systems and enhancing their practical performance under dynamic channel and network conditions. Specifically, we propose the Wireless AutoResearch Agent (WARA), a closed-loop multi-agent system for automated wireless optimization research. Given only an initial research topic as input, WARA decomposes the research workflow into three phases: 1) research gap identification and problem proposal, 2) wireless optimization problem modeling, solution algorithm design, and experimentation, and 3) construction of research deliverables. For each phase, we design artifact-mediated control mechanisms, where declared upstream artifacts are consumed as inputs to produce structured outputs for downstream use. A controller-managed gate is then applied in each phase to validate the generated artifacts and ensure consistency among models, algorithms, experiments, and claims. When an artifact fails validation, only the corresponding artifact needs to be repaired, rather than restarting the entire workflow. Furthermore, we present a representative wireless resource allocation case study to demonstrate how WARA converts an initial topic into a complete research package with executable evidence and a synthesized technical manuscript. We also design a structured LLM-based ScoringAgent to evaluate manuscript-level research validity and optimization research maturity, and the comparative evaluations show that WARA substantially outperforms one-shot LLM generation while approaching the quality profile of recently accepted peer-reviewed technical papers. These results demonstrate that closed-loop artifact control is a promising path toward end-to-end LLM-assisted wireless optimization research. The source code is available at \url{https://github.com/guoyuan-dotcom/WARA_CUHKSZ}.
	\end{abstract}
	

	\section{Introduction}
	
	The remarkable success of large artificial intelligence models, particularly large language models (LLMs), has reshaped expectations regarding the role of artificial intelligence (AI) in scientific and engineering research. Beyond their established use in text and code generation, LLMs are increasingly being embedded into agentic systems that integrate language-based reasoning with external tools, executable environments, reflection mechanisms, and multi-agent coordination ~\cite{yao2023react,schick2023toolformer,shinn2023reflexion,wu2023autogen}. In this context, LLM-based agentic systems offer a promising path toward autonomous research exploration, enabling the generation of innovative research ideas and their progressive refinement into validated research outcomes through iterative reasoning, execution, and verification.
	
	A typical research workflow in science and engineering can generally be divided into three phases: $1)$ research gap and problem identification, where a broad topic or motivation is used to guide a comprehensive literature review, identify limitations in prior work, and formulate a concrete research problem; $2)$ problem solving and experimentation, where the problem is developed into a technical model, solved using suitable algorithms or methods, and evaluated through experiments; and $3)$ construction of deliverables, where the results are verified, refined, and organized into research papers through iterative research-review cycles. 
	
	Prior work has shown that LLMs and LLM-based agents can assist research in each of these phases separately. First, for the research gap and problem identification phase, the authors in \cite{si2024ideas} exploited LLMs to generate research ideas and evaluated their quality against human-generated ideas through expert review. ResearchAgent \cite{baek2024researchagent} and Scideator \cite{radensky2024scideator} further used existing papers to support research problem generation and human–LLM ideation, while the AI co-scientist system \cite{gottweis2025aicoscientist} employed multiple agents to generate and refine scientific hypotheses. Next, for the problem solving and experimentation phase, several studies have investigated local design–execution loops, where LLM agents write code, run experiments, inspect outputs, and revise their solutions \cite{huang2023mlagentbench,chan2024mlebench,jiang2025aide,ramamonjison2023nl4opt,huang2024mamo,huang2025orlm,ahmaditeshnizi2024optimus,chen2025optichat}. Specifically, MLAgentBench \cite{huang2023mlagentbench}, MLEbench \cite{chan2024mlebench}, and AIDE \cite{jiang2025aide} evaluated these capabilities in machine-learning experimentation and engineering settings. \cite{ramamonjison2023nl4opt,huang2024mamo,huang2025orlm,ahmaditeshnizi2024optimus,chen2025optichat} studied LLM-assisted optimization scenarios, where LLMs transform natural-language problem descriptions into mathematical optimization formulations and then design algorithms or solution methods for the formulated problems. Furthermore, for the construction of deliverables phase, data-to-paper \cite{ifargan2024datatopaper} and CycleResearcher \cite{weng2024cycleresearcher} studied how LLMs can organize technical results into complete research papers through iterative research-review cycles. Relatedly, \cite{liang2023llmfeedback,zhu2025deepreview,starace2025paperbench} investigated LLM-based review and evaluation methods to support scientific feedback, paper review, and paper-level assessment.

	Different from the above works that focus on specific phases of the research process, another line of research takes a further step by connecting these phases into closed-loop research workflows, leading to recent advances in end-to-end closed-loop autoresearch. In particular, autoresearch has attracted growing interest in AI/machine learning (ML) and computational research, where algorithm design and experimentation can be performed online by LLM agents through code generation and tool use. For example, AI Scientist is an early end-to-end AI/ML autoresearch system that generates complete papers from research ideas, but it largely relies on single-agent reasoning and has limited ability to recover from execution failures \cite{lu2024aiscientist}. AI Scientist-v2 advances this direction through agentic tree search and experiment management, although each run is still largely treated as an independent research attempt \cite{yamada2025aiscientistv2}. AI-Researcher develops a multi-agent system for autonomous AI research and introduces Scientist-Bench to evaluate the quality of generated research outputs \cite{tang2025airesearcher}. AutoResearchClaw \cite{autoresearchclaw2026} further formulates autonomous research as an iterative process by incorporating failure recovery, result verification, human–AI collaboration, and experience accumulation across runs. Beyond AI/ML, a recent work \cite{ghareeb2026multi} proposed Robin, an autoresearch framework for science, particularly biology, where LLM-based multi-agent systems are used for background research, hypothesis generation, and data analysis, while human-based laboratory experimentation is integrated as one phase of the overall loop.

	Motivated by recent advances in autoresearch for science and engineering, this paper investigates closed-loop research automation for wireless communications, with a particular focus on wireless resource allocation and optimization. Resource allocation is a cornerstone of wireless communication research, playing a central role in both characterizing fundamental performance limits and enhancing practical system performance under dynamic channel and network conditions\footnote{A survey of major IEEE wireless communications venues from 2010 to 2025 reveals that optimization-related research consistently constitutes a substantial portion of published papers, with the proportion typically ranging from 40\% to 70\% across different venues and years.}. On the one hand, many fundamental problems in wireless communications can be formulated as optimization problems. For example, the characterization of information-theoretic limits in communication systems such as multi-input multi-output (MIMO) communications \cite{1266912PAULRAJ}, massive MIMO \cite{larsson2014massivemimo}, reconfigurable intelligent surface (RIS)-assisted networks \cite{wu2019intelligent}, movable-antenna systems \cite{zhu2023movable}, and unmanned aerial vehicle (UAV)-enabled communications \cite{Zeng7470933}, integrated sensing and communications (ISAC) \cite{Liu9737357}, and simultaneous wireless information and power transfer (SWIPT) \cite{Xu6860253}, can often be cast as rate maximization problems over transmit covariance matrices, power allocation policies, bandwidth allocation strategies, and other communication resources \cite{goldsmith2005wireless,tse2005fundamentals}. On the other hand, future sixth-generation (6G) networks are expected to operate in highly dynamic environments with massive antenna arrays, dense deployments, heterogeneous nodes, integrated sensing and computing functionalities, and rapidly varying service requirements. In such systems, adaptive resource allocation is essential for improving communication efficiency, mitigating interference, and coordinating communication, sensing, and computing resources across space and time. Importantly, wireless optimization research exhibits a highly structured workflow that is particularly amenable to automation. Given a communication scenario, researchers typically formulate a mathematical optimization problem based on physical channel models, communication-performance metrics, and resource constraints~\cite{tse2005fundamentals,goldsmith2005wireless,palomar2010convex,chiang2007layering,luo1664998,liu10636212}. The resulting problem is then solved using analytical or numerical methods, followed by simulation-based evaluation under representative network settings. This workflow naturally provides executable verification mechanisms, where optimization algorithms and simulation environments can be repeatedly invoked to validate technical claims. Such a process of modeling assumptions, optimization formulations, algorithm design, experimental evaluation, and performance claims makes wireless optimization research an ideal test-bed for closed-loop autoresearch systems.

	In the literature, recent studies have begun exploring the use of LLMs and LLM agents in wireless communications. One line of work focuses on LLM-assisted wireless optimization and network control, including power control, adaptive resource allocation, optimization reformulation, and LLM-powered xApps for O-RAN management~\cite{qiu2024llmwirelessdesign,zhou2024llmpowercontrol,lee2026llmresource,noh2025adaptive,peng2025llmoptira,wu2025llmxapp}. Another line of work develops wireless-domain LLM agent frameworks and benchmarks. Examples include LLM-enhanced multi-agent systems for 6G communications \cite{du2023llm6gmas}, WirelessAgent for intelligent network management \cite{tong2025wirelessagent}, AgentRAN for autonomous O-RAN control \cite{elkael2025agentran}, ComAgent for intent-driven wireless optimization and simulation generation \cite{li2026comagent}, WirelessAgent++ for workflow discovery, and WirelessBench for wireless-agent evaluation \cite{tong2026wirelessagentpp,tong2026wirelessbench}. While these efforts demonstrate the potential of LLMs for wireless optimization and network intelligence, they primarily address isolated tasks such as network control, optimization formulation, workflow generation, or benchmark evaluation. To the best of our knowledge, no existing work has investigated the end-to-end automation of wireless optimization research, spanning research problem discovery, mathematical modeling, algorithm design, experimentation, result validation, and manuscript generation within a unified closed-loop framework.
	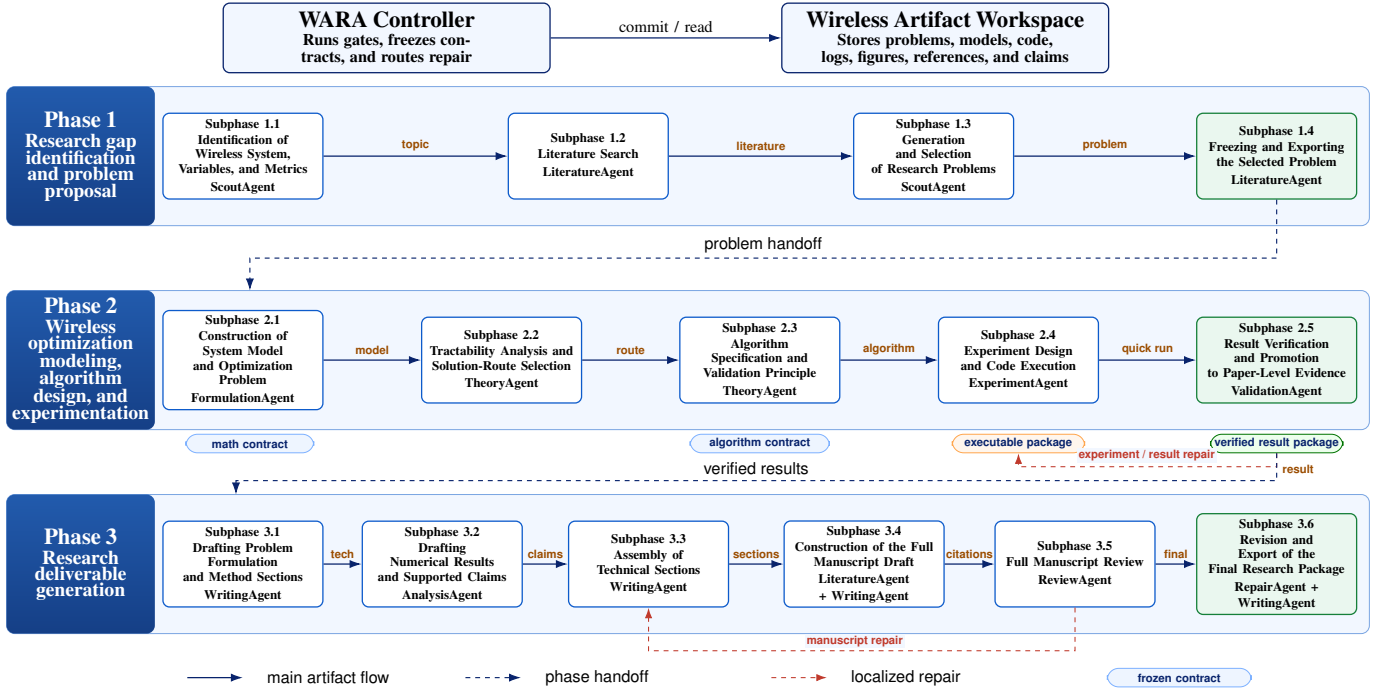
\begin{figure*}[!t]
		\centering
		\resizebox{\textwidth}{!}{%
			\begin{tikzpicture}[
				x=1cm,
				y=1cm,
				font=\small,
				flow/.style={-{Latex[length=2.6mm,width=1.8mm]}, draw=waraNavy, line width=0.78pt},
				handoff/.style={-{Latex[length=2.6mm,width=1.8mm]}, draw=waraNavy, dashed, dash pattern=on 3pt off 3pt, line width=0.70pt},
				repair/.style={-{Latex[length=2.3mm,width=1.6mm]}, draw=waraRepair, dashed, dash pattern=on 3pt off 3pt, line width=0.65pt},
				lane/.style={draw=waraBlue!55, rounded corners=5pt, line width=0.56pt, fill=waraLane},
				phasebox/.style={draw=waraBlue, rounded corners=4pt, line width=0.80pt, fill=white, align=center, text width=3.10cm, minimum width=3.45cm, minimum height=1.85cm, inner sep=3pt},
				frozenstep/.style={draw=waraGreenLine, rounded corners=4pt, line width=0.80pt, fill=waraGreen, align=center, text width=3.10cm, minimum width=3.45cm, minimum height=1.85cm, inner sep=3pt},
				topbox/.style={draw=waraNavy, rounded corners=4pt, line width=0.80pt, fill=waraLane, align=center, text width=6.80cm, minimum width=7.05cm, minimum height=1.5cm, inner sep=2pt},
				tag/.style={draw=waraTagLine, rounded corners=7pt, line width=0.50pt, fill=waraTag, align=center, minimum height=0.40cm, inner sep=1.5pt},
				taggreen/.style={draw=green!45!black, rounded corners=7pt, line width=0.50pt, fill=green!7, align=center, minimum height=0.40cm, inner sep=1.5pt},
				tagorange/.style={draw=orange!70, rounded corners=7pt, line width=0.50pt, fill=orange!7, align=center, minimum height=0.40cm, inner sep=1.5pt},
				side/.style={rounded corners=5pt, align=center, text=white},
				gatelabel/.style={font=\sffamily\bfseries\scriptsize, text=waraOrange},
				repairlabel/.style={font=\sffamily\bfseries\scriptsize, text=waraRepair, fill=waraLane, inner sep=1pt},
				tagtext/.style={font=\sffamily\bfseries\scriptsize, text=waraNavy}
				]
				
				\node[topbox] (controller) at (8.70,14)
				{\Large\bfseries WARA Controller\\[0.5mm]
					\normalsize Runs gates, freezes contracts, and routes repair};
				
				\node[topbox] (workspace) at (20.75,14)
				{\Large\bfseries Wireless Artifact Workspace\\[0.5mm]
					\normalsize Stores problems, models, code, logs, figures, references, and claims};
				
				\draw[flow] (controller.east) -- node[above, font=\normalsize] {commit / read} (workspace.west);
				
				\draw[lane] (0.50,9.95) rectangle (29.9,12.95);
				\draw[lane] (0.50,5.55) rectangle (29.9,8.55);
				\draw[lane] (0.50,1.15) rectangle (29.9,4.15);
				
				\shade[rounded corners=5pt, top color=waraSideTop, bottom color=waraSideBottom]
				(0.50,9.95) rectangle (3.70,12.95);
				\shade[rounded corners=5pt, top color=waraSideTop, bottom color=waraSideBottom]
				(0.50,5.55) rectangle (3.70,8.55);
				\shade[rounded corners=5pt, top color=waraSideTop, bottom color=waraSideBottom]
				(0.50,1.15) rectangle (3.70,4.15);
				
				\node[side, text width=3.05cm] at (2.10,11.45)
				{\Large\bfseries Phase 1\\[0.8mm]\large\bfseries Research gap identification and problem proposal};
				
				\node[side, text width=3.05cm] at (2.10,7.05)
				{\Large\bfseries Phase 2\\[0.8mm]\large\bfseries Wireless optimization modeling, algorithm design, and experimentation};
				
				\node[side, text width=3.05cm] at (2.10,2.65)
				{\Large\bfseries Phase 3\\[0.8mm]\large\bfseries Research deliverable generation};
				
				\node[phasebox] (p11) at (5.6,11.45)
				{\footnotesize\bfseries Subphase 1.1\\[0.3mm]
					 Identification of Wireless System,\\Variables, and Metrics\\[0.3mm]
					{\footnotesize ScoutAgent}};
				
				\node[phasebox] (p12) at (13.05,11.45)
				{\footnotesize\bfseries Subphase 1.2\\[0.3mm]
					 Literature Search\\[0.3mm]
					{\footnotesize LiteratureAgent}};
				
				\node[phasebox] (p13) at (20.50,11.45)
				{\footnotesize\bfseries Subphase 1.3\\[0.3mm]
					 Generation and Selection\\of Research Problems\\[0.3mm]
					{\footnotesize ScoutAgent}};
				
				\node[frozenstep] (p14) at (27.9,11.45)
				{\footnotesize\bfseries Subphase 1.4\\[0.3mm]
					 Freezing and Exporting\\the Selected Problem\\[0.3mm]
					{\footnotesize LiteratureAgent}};
				
				\draw[flow] (p11.east) -- node[above, gatelabel] {topic} (p12.west);
				\draw[flow] (p12.east) -- node[above, gatelabel] {literature} (p13.west);
				\draw[flow] (p13.east) -- node[above, gatelabel] {problem} (p14.west);
				
				\draw[handoff] (p14.south) -- (27.9,9.25)
				-- node[above, font=\sffamily\normalsize] {problem handoff}
				(5.75,9.25) -- (5.75,8.55);
				
				\node[phasebox] (p21) at (5.6,7.05)
				{\footnotesize\bfseries Subphase 2.1\\[0.3mm]
					\footnotesize Construction of System Model\\and Optimization Problem\\[0.3mm]
					{\footnotesize FormulationAgent}};
				
				\node[phasebox] (p22) at (11.18,7.05)
				{\footnotesize\bfseries Subphase 2.2\\[0.3mm]
					\footnotesize Tractability Analysis and\\Solution-Route Selection\\[0.3mm]
					{\footnotesize TheoryAgent}};
				
				\node[phasebox] (p23) at (16.75,7.05)
				{\footnotesize\bfseries Subphase 2.3\\[0.3mm]
					\footnotesize Algorithm Specification and\\Validation Principle\\[0.3mm]
					{\footnotesize TheoryAgent}};
				
				\node[phasebox] (p24) at (22.33,7.05)
				{\footnotesize\bfseries Subphase 2.4\\[0.3mm]
					\footnotesize Experiment Design\\and Code Execution\\[0.3mm]
					{\footnotesize ExperimentAgent}};
				
				\node[frozenstep] (p25) at (27.9,7.05)
				{\footnotesize\bfseries Subphase 2.5\\[0.3mm]
					\footnotesize Result Verification and Promotion\\to Paper-Level Evidence\\[0.3mm]
					{\footnotesize ValidationAgent}};
				
				\draw[flow] (p21.east) -- node[above, gatelabel] {model} (p22.west);
				\draw[flow] (p22.east) -- node[above, gatelabel] {route} (p23.west);
				\draw[flow] (p23.east) -- node[above, gatelabel] {algorithm} (p24.west);
				\draw[flow] (p24.east) -- node[above, gatelabel] {quick run} (p25.west);
				
				\node[tag, tagtext, text width=2.65cm] at (5.75,5.25) {math contract};
				\node[tag, tagtext, text width=2.90cm] at (16.75,5.25) {algorithm contract};
				\node[tagorange, tagtext, text width=2.75cm] (xpack) at (22.33,5.25) {executable package};
				\node[taggreen, tagtext, text width=2.75cm] (evtag) at (27.9,5.25) {verified result package};
				
				\draw[repair] (evtag.south) -- (27.9,4.75) -- (22.33,4.75) -- (xpack.south);
				\node[repairlabel] at (25.10,5.00) {experiment / result repair};
				
				\draw[handoff] (evtag.south) -- node[right, yshift=-1pt, gatelabel] {result}
				(27.9,4.45) -- node[above, font=\sffamily\normalsize] {verified results}
				(5.45,4.45) -- (5.45,4.15);
				
				\node[phasebox] (p31) at (5.6,2.65)
				{\footnotesize\bfseries Subphase 3.1\\[0.3mm]
					\footnotesize Drafting Problem Formulation\\and Method Sections\\[0.3mm]
					{\footnotesize WritingAgent}};
				
				\node[phasebox] (p32) at (9.90,2.65)
				{\footnotesize\bfseries Subphase 3.2\\[0.3mm]
					\footnotesize Drafting Numerical Results\\and Supported Claims\\[0.3mm]
					{\footnotesize AnalysisAgent}};
				
				\node[phasebox] (p33) at (14.35,2.65)
				{\footnotesize\bfseries Subphase 3.3\\[0.3mm]
					\footnotesize Assembly of\\Technical Sections\\[0.3mm]
					{\footnotesize WritingAgent}};
				
				\node[phasebox] (p34) at (19,2.65)
				{\footnotesize\bfseries Subphase 3.4\\[0.3mm]
					\footnotesize Construction of the Full\\Manuscript Draft\\[0.3mm]
					{\footnotesize LiteratureAgent + WritingAgent}};
				
				\node[phasebox] (p35) at (23.55,2.65)
				{\footnotesize\bfseries Subphase 3.5\\[0.3mm]
					\footnotesize Full Manuscript Review\\[0.3mm]
					{\footnotesize ReviewAgent}};
				
				\node[frozenstep] (p36) at (27.9,2.65)
				{\footnotesize\bfseries Subphase 3.6\\[0.3mm]
					\footnotesize Revision and Export of the\\Final Research Package\\[0.3mm]
					{\footnotesize RepairAgent + WritingAgent}};
				
				\draw[flow] (p31.east) -- node[above, gatelabel] {tech} (p32.west);
				\draw[flow] (p32.east) -- node[above, gatelabel] {claims} (p33.west);
				\draw[flow] (p33.east) -- node[above, gatelabel] {sections} (p34.west);
				\draw[flow] (p34.east) -- node[above, gatelabel] {citations} (p35.west);
				\draw[flow] (p35.east) -- node[above, gatelabel] {final} (p36.west);
				
				\draw[repair] (p35.south) -- (23.55,0.78) -- (14.35,0.78) -- (p33.south);
				\node[repairlabel] at (18.80,0.98) {manuscript repair};
				
				\draw[flow] (4.4,0.20) -- +(1.2,0);
				\node[anchor=west, font=\sffamily\normalsize] at (6.0,0.20) {main artifact flow};
				
				\draw[handoff] (10.4,0.20) -- +(1.2,0);
				\node[anchor=west, font=\sffamily\normalsize] at (12.0,0.20) {phase handoff};
				
				\draw[repair] (17.0,0.20) -- +(1.2,0);
				\node[anchor=west, font=\sffamily\normalsize] at (18.6,0.20) {localized repair};
				
				\node[tag, tagtext, text width=3.0cm] at (25.8,0.20) {frozen contract};
				
			\end{tikzpicture}
		}
		\caption{WARA workflow for wireless optimization autoresearch. Each subphase is executed by one or more role-specialized agents responsible for specific tasks, as summarized in Table \ref{tab:wara_agents}.}
		\label{fig:wara_pipeline}
	\end{figure*}

	To bridge this gap, we propose the Wireless AutoResearch Agent (WARA), an end-to-end autoresearch framework for wireless optimization. Rather than generating individual artifacts in isolation, WARA aims to transform an initial wireless research topic into a technically consistent, evidence-supported, and reviewable research package, where modeling assumptions, optimization formulations, algorithmic designs, executable experiments, and reported claims remain coherent throughout the entire research lifecycle. Specifically, WARA organizes the research process into three phases: 1) research gap identification and problem proposal, 2) wireless optimization modeling and experimentation, and 3) research deliverable generation. For each phase, we design artifact-mediated control mechanisms, where declared upstream artifacts are consumed as inputs to produce structured outputs for downstream use. A controller-managed gate is then applied in each phase to validate the generated artifacts and ensure consistency among models, algorithms, experiments, and claims. When an artifact fails validation, only the corresponding artifact needs to be repaired, rather than restarting the entire workflow. 
	
	Furthermore, we present a representative wireless resource allocation case study to demonstrate how WARA transforms an initial research topic into a complete set of research deliverables, including executable experimental evidence and a synthesized technical manuscript. To evaluate the generated manuscripts, we further design a structured LLM-based ScoringAgent that applies fixed reviewer-style rubrics to assess manuscript-level research validity and optimization research maturity. Comparative evaluations show that WARA substantially outperforms one-shot LLM generation under the same topic set and backbone model, while approaching the quality profile of recently accepted peer-reviewed technical papers. These results highlight the effectiveness of closed-loop artifact-mediated control and suggest a promising pathway toward end-to-end LLM-assisted wireless optimization research.
	
	The remainder of this paper is organized as follows. Section~\ref{sec:architecture} proposes the WARA framework and details the three research phases. Section~\ref{sec:case_studies} presents a representative wireless optimization case study and illustrates the complete artifact trajectory generated by WARA. Section~\ref{sec:results} presents comparative evaluations against one-shot LLM-generated manuscripts and recently accepted peer-reviewed papers. Finally, Section~\ref{sec:conclusion} concludes the paper and discusses future research directions for wireless autoresearch.

	\section{WARA Framework}
	\label{sec:architecture}
	In this section, we present the WARA framework for closed-loop wireless optimization research. We first introduce the overall architecture and workflow of WARA. We then describe the implementation details of each phase, followed by the artifact-mediated control mechanism consisting of gate validation, interface freezing, and localized repair.
	
	\subsection{Overall Workflow}
	\label{subsec:problem_setting}
	Following the standard workflow of wireless optimization research, WARA is designed to automatically construct a complete research package from an initial wireless research topic under a finite computational budget. As illustrated in Fig. \ref{fig:wara_pipeline}, WARA organizes the research process into three sequential phases coordinated by multiple role-specialized LLM agents, i.e.
	\begin{enumerate}
		\item[1)]	Research Gap Identification and Problem Proposal;
		\item[2)]	Wireless Optimization Modeling, Algorithm Design, and Experimentation;
		\item[3)]	Research Deliverable Generation.
	\end{enumerate}
	
	In particular, the first phase aims to transform a broad research topic into a concrete and technically grounded research problem. To achieve this, it is further divided into four subphases that successively generate a research frame, collect grounding evidence from the literature, produce and select candidate research directions, and export a frozen problem handoff. Given the selected research problem, the second phase develops the technical solution through five subphases. Specifically, it constructs the system model, formulates the optimization problem, analyzes problem tractability and potential reformulations, designs the solution algorithm and validation principles, and generates executable experimental workflows. Through iterative execution, quick validation, and result verification, this phase ultimately produces validated numerical results and paper-ready figures supporting the proposed technical approach. Finally, the third phase transforms the generated technical artifacts into research deliverables. Specialized agents are responsible for drafting and assembling manuscript components, including the introduction, system model, methodology, experimental evaluation, and conclusion sections. These components are subsequently integrated into a complete manuscript package. In parallel, a review agent evaluates the technical consistency, result support, and overall manuscript quality, and requests revisions when deficiencies are identified.
	
	For the purpose of clarity, we now define several concepts used throughout the WARA workflow as follows,
	\begin{itemize}
		\item \textbf{Artifact} is a stored research object produced or used during a WARA run. It records content generated by an agent, an execution tool, or a validation step. The workspace saves artifacts as persistent files so that later steps can read, check, and reuse them.
		
		\item \textbf{Role-specialized agents} are task-specific LLM roles used to produce or revise WARA artifacts. Each agent uses its role prompt and the relevant input artifacts to generate the required output, which is then stored in the workspace and checked by the corresponding gate. The main agents, their research roles, artifact interfaces, and execution or verification mechanisms are summarized in Table~\ref{tab:wara_agents}.
		
		\item \textbf{Gate} is a check performed before an artifact, contract, or handoff is used by later steps. It produces a machine-readable report with the pass/fail result, detected problems, and repair suggestions. The gate checks the required properties of the current subphase, including completeness, contract consistency, executable behavior, evidence validity, citation support, and manuscript consistency.
		
		\item \textbf{Contract} is a structured artifact that records technical decisions to be preserved at later steps. A contract fixes decisions made in an accepted artifact, including the problem direction, modeling assumptions, notation, variables, objective, constraints, solution route, solver requirements, experiment scope, or claim boundary. After the corresponding gate passes, the contract is frozen and later agents must follow it when generating, checking, or revising artifacts.
		
		\item \textbf{Handoff} is the information package passed from one phase to the next after the current phase has been checked. It collects the accepted artifacts and frozen contracts required by the next phase. A handoff records what the next phase uses as its fixed starting point, including the selected direction, modeling requirements, validation plan, verified evidence, and supported claims.
		
		\item \textbf{Controller} is the module that coordinates a WARA run. It calls agents, records their outputs, tracks dependencies among artifacts, applies gates, freezes accepted contracts and handoffs, and sends failed outputs back for revision. The controller records these actions in run manifests rather than rewriting the research content.
	\end{itemize}

	In the following three subsections, we describe the execution of the three phases by detailing their corresponding subphases.
	
	\begin{table*}[!t]
		\centering
		\caption{Main Role-Specialized Agents in WARA.}
		\label{tab:wara_agents}
		\scriptsize
		\setlength{\tabcolsep}{3pt}
		\renewcommand{\arraystretch}{1.18}
		\begin{tabular}{p{0.13\linewidth} p{0.30\linewidth} p{0.31\linewidth} p{0.18\linewidth}}
			\hline
			\textbf{Agent} & \textbf{Research role} & \textbf{Input / output artifacts} & \textbf{Execution / verification interface} \\
			\hline
			
			ScoutAgent 
			& Frames the initial wireless topic, defines the research scope, compares candidate directions, and selects a bounded problem direction for downstream modeling. 
			& Reads the user topic and available literature signals. Writes the research frame, candidate directions, selected direction, and Phase-1 handoff. 
			& Schema validation; scope and feasibility checks. \\
			\hline
			
			LiteratureAgent 
			& Grounds the selected direction in wireless literature and identifies prior models, baselines, assumptions, and citation needs. 
			& Reads the research frame, selected direction, and search plan. Writes the evidence pack, reference bank, and citation--claim map. 
			& Literature retrieval; metadata verification; citation consistency checks. \\
			\hline
			
			FormulationAgent 
			& Converts the selected direction into a wireless system model and an original optimization formulation. 
			& Reads the Phase-1 handoff and evidence pack. Writes the system model, assumptions, variables, objective, constraints, and mathematical contract. 
			& Symbol-role audit; equation consistency checks; \LaTeX{} validation. \\
			\hline
			
			TheoryAgent 
			& Analyzes tractability, identifies the reformulation route, and develops the algorithmic solution scope. 
			& Reads the frozen mathematical contract, system model, and problem formulation. Writes the tractability analysis, reformulation path, algorithm contract, and algorithm description. 
			& Contract compliance checks; convexity and assumption audit. \\
			\hline
			
			ExperimentAgent 
			& Builds and executes the experiment package according to the frozen mathematical and algorithmic contracts. 
			& Reads the mathematical contract, algorithm contract, algorithm description, and experiment blueprint. Writes the experiment design, Python/CVXPY code, benchmark implementations, execution logs, solver outputs, and candidate figures. 
			& Code execution; solver validation; simulation logs; benchmark checks; plot/data generation. \\
			\hline
			
			ValidationAgent 
			& Checks experiment outputs and promotes valid numerical evidence for downstream analysis and writing. 
			& Reads the experiment report, figure data, validation logs, and benchmark outputs. Writes the evidence contract, verified figure evidence, benchmark definitions, and evidence-readiness summary. 
			& CSV/log inspection; figure audit; benchmark consistency checks; metric and claim-evidence checks. \\
			\hline
			
			AnalysisAgent 
			& Interprets verified numerical evidence and maps supported trends to research claims. 
			& Reads the evidence contract and verified figure evidence. Writes the numerical-results text, claim--evidence map, and result interpretation notes. 
			& Claim--evidence checks; trend consistency audit. \\
			\hline
			
			WritingAgent 
			& Assembles manuscript text from frozen contracts, verified evidence, and curated citations. 
			& Reads the mathematical contract, algorithm contract, numerical-results text, figure evidence, and reference bank. Writes technical sections, full-paper preview, and revised manuscript text. 
			& \LaTeX{} compilation; bibliography validation; style and notation checks. \\
			\hline
			
			ReviewAgent 
			& Audits the artifact chain and final draft for technical consistency, evidence support, citation adequacy, and manuscript readiness. 
			& Reads the full-paper preview, citation--claim map, experiment report, figure evidence, and frozen contracts. Writes the review report, critical issues, and routing decision. 
			& Cross-artifact audit; evidence and citation checks; PDF/log inspection. \\
			\hline
			
			RepairAgent 
			& Applies scoped revisions to the responsible artifact according to gate or review feedback. 
			& Reads the target artifact, error report, and relevant frozen contract. Writes the repaired artifact, repair log, and gate rerun result. 
			& Localized repair; regression checks; gate rerun. \\
			\hline
		\end{tabular}
	\end{table*}
	
	\subsection{Phase 1: Research Gap Identification and Problem Proposal}
	\label{subsec:phase1_gap_problem}

	Phase 1 addresses the transition from a broad wireless research topic to a well-defined wireless optimization problem. Its design is inspired by topic-grounding mechanisms in automated research workflows \cite{autoresearchclaw2026}, while being specialized for wireless optimization by jointly considering the communication scenario, controllable resources, performance objectives, and supporting literature. This phase consists of four subphases that progressively generate, evaluate, select, and formalize a research problem for the technical development in the next phase.
	\subsubsection{Subphase 1.1:  Identification of Wireless System, Variables, and Metrics}
	Given an initial wireless research topic, Subphase 1.1 transforms it into a structured wireless optimization frame. Specifically, the controller first validates the topic against a wireless-domain taxonomy\footnote{The taxonomy covers common wireless tasks, network settings, radio technologies, optimization variables, modeling assumptions, solver families, and performance metrics. The topic may include ISAC, SWIPT, cell-free networks, RIS,  beamforming, and power allocation. Model assumptions, optimization techniques, and objectives may include imperfect channel state information (CSI), successive convex approximation (SCA), semidefinite relaxation (SDR), sum rate, harvested power, and Cram\'er--Rao bound (CRB). } to filter out unrelated systems, objectives, variables, and metrics. The ScoutAgent then identifies the corresponding wireless scenario, key physical tradeoffs, optimization variables, candidate objectives, major constraints, and evaluation metrics. The resulting research frame serves as the foundation for subsequent literature search and problem formulation.
	
	\subsubsection{Subphase 1.2: Literature Search}
	Using the research frame generated in Subphase 1.1, the LiteratureAgent constructs search queries based on the identified communication system, optimization variables, objectives, performance metrics, and potential solution directions. Literature retrieval is then conducted through multiple academic sources, including Crossref, Semantic Scholar, OpenAlex, arXiv, IEEE Xplore, and Google Scholar. If the retrieved literature is insufficient, the search is expanded using a local repository of seminal papers and additional query refinements. The output of this subphase is an evidence pack containing related system models, assumptions, baselines, performance metrics, references, and citation requirements, which provides the knowledge foundation for candidate-problem generation and later validation.
	
	\subsubsection{Subphase 1.3: Generation and Selection of Research Problems}
	Based on the research frame and evidence pack, the ScoutAgent generates multiple candidate wireless optimization problems and recommends one for further development. Rather than directly accepting the recommendation, the controller evaluates each candidate according to predefined criteria, including field completeness, consistency with the wireless-domain taxonomy, logical coherence of the underlying mechanisms, and suitability for downstream optimization modeling and experimentation in Phase 2. A candidate is accepted only if it is sufficiently specific, optimization-oriented, technically grounded, and actionable. For traceability, the selected problem is stored together with the alternative candidates and the rationale underlying the selection decision.
	
	\subsubsection{Subphase 1.4:Freezing and Exporting the Selected Problem}
	Subphase 1.4 converts the selected problem into a formal handoff artifact for Phase 2. Given the selected problem, supporting references, and candidate-selection record, the LiteratureAgent performs a focused literature search to further verify topic-specific grounding and ensure adequate evidential support. If the retrieved references are insufficient, the search scope is iteratively expanded until the required coverage is achieved. The controller then assembles a structured problem contract containing the selected problem definition, relevant references, and modeling instructions required for optimization formulation, algorithm design, experimentation, and validation in Phase 2. Once the handoff artifact passes the validation gate, it is frozen and delivered to Phase 2. Otherwise, it is routed back to the literature-search or problem-selection subphases for targeted repair and refinement.

	\subsection{Phase 2: Wireless Optimization Modeling, Solving, and Experimentation}
	\label{subsec:phase2_modeling_experimentation}
	
	With the Phase-1 handoff frozen, Phase 2 transforms the selected research problem into a wireless system model and optimization formulation, derives a solution route consistent with the formulation, and implements the resulting method in executable experiment code. The following subphases describe this process from model construction to the generation of verified numerical evidence.

	\subsubsection{Subphase 2.1: Construction of System Model and Optimization Problem}
	Using the frozen problem handoff from Phase 1 as input, FormulationAgent invokes an LLM to generate the wireless system model and the corresponding optimization problem. The resulting artifact specifies the system setup, including signal and channel models, optimization variables, objective function, constraints, and performance metrics. The controller then verifies whether the formulation is complete, physically meaningful, and consistent with the selected research problem. If the formulation gate fails, the detected issues are returned to FormulationAgent for repair within a bounded repair budget. If the gate passes, the system model and optimization formulation are frozen as the mathematical contract. If no acceptable formulation is obtained after the allowed repair attempts, the run is terminated and the formulation errors are recorded in the repair log.

	\subsubsection{Subphase 2.2: Tractability Analysis and Solution-Route Selection}
	Using the frozen mathematical contract from Subphase 2.1, the controller prepares a tractability-analysis prompt and passes it to TheoryAgent. TheoryAgent then invokes an LLM to determine whether the optimization problem is directly tractable or requires reformulation. The analysis examines convexity, conic representability, and major sources of difficulty, such as nonconvex coupling, discrete variables, rank constraints, fractional structures, and uncertainty. If the problem is directly tractable, a convex or conic optimization solver is selected. Otherwise, an appropriate reformulation or approximation strategy is chosen, such as SDR, SCA, bisection, alternating optimization, robust approximation, mixed-integer optimization, or a problem-specific heuristic.

	\subsubsection{Subphase 2.3: Algorithm Specification and Validation Principle}
	Based on the mathematical contract and the selected solution route, TheoryAgent invokes an LLM to generate an implementable algorithm specification. The resulting artifact describes the variable-update rules, feasibility checks, stopping criteria, and baseline-comparison methodology. The controller then verifies consistency between the algorithm and the frozen optimization formulation, and checks whether the associated convergence, complexity, and optimality discussions are supported by the chosen solution route. If the theory gate passes, the algorithm specification and validation principles are accepted, and a claim map is created to track theoretical and empirical claims throughout subsequent stages. Otherwise, the identified issues are returned to TheoryAgent for targeted repair.

	\subsubsection{Subphase 2.4: Experiment Design and Code Execution}
	Using the mathematical contract, algorithm contract, and claim map as inputs, the controller first constructs a structured experiment-design specification, including comparison methods, parameter sweeps, performance metrics, and figure objectives. ExperimentAgent then invokes an LLM to generate a concrete experiment design. The controller verifies that the selected metrics, sweep variables, and benchmark methods are consistent with the frozen optimization problem. Once the design passes validation, ExperimentAgent generates Python experiment code.
	
	The controller subsequently inserts the generated code into a fixed validation harness\footnote{A validation harness is controller-owned Python code to standardize how the generated experiment code is loaded, executed, logged, and checked. It
		provides the fixed execution wrapper around the problem-specific experiment
		code and implements the feedback and verification layer required for reliable
		agent execution \cite{zhong2026aiharness}.} and executes it locally. The harness performs interface validation, code execution, log collection, result-table generation, and output verification. It also checks whether the implementation faithfully follows the algorithm specification. For solver-based approaches, the implementation must invoke the designated solver rather than replacing it with a proxy implementation. If validation succeeds, an executable experiment package is produced. Otherwise, the code, execution logs, validation report, and frozen contracts are returned to ExperimentAgent for bounded repair.

	\subsubsection{Subphase 2.5: Result Verification and Promotion to Paper-Level Evidence}
	Subphase 2.5 transforms the executable experiment package into verified numerical evidence. ValidationAgent examines the approved experiment design, benchmark plan, execution logs, result tables, and candidate figures to determine whether the generated results support the intended wireless communication claims. The validation process checks data provenance, benchmark consistency, metric validity, and figure readiness. Specifically, plotted curves must originate from saved result tables, compared methods must match the approved design, metrics must correspond to valid wireless-performance measures, and figures must satisfy manuscript-quality requirements.
	
	After initial validation, a three-level experiment-expansion strategy is employed. The first level performs a scout sweep to identify the intended trend using a small number of representative parameter settings. The second level conducts a medium-scale sweep to verify trend stability using additional parameter values and random seeds. The final level performs a paper-scale sweep using denser parameter grids and larger numbers of random seeds to generate publication-quality figures. Throughout all stages, the benchmark family, metric family, and target claims remain fixed according to the approved experiment design. If inconsistencies are detected, the issue is routed back to experiment-design repair. Once the final validation gate passes, the verified result package is frozen and handed over to Phase 3.

	\subsection{Phase 3: Research Deliverable Generation}
	\label{subsec:phase3_writing_export}
	
	Phase 3 transforms the validated outputs of Phase 2 into a complete research package. It generates the manuscript, validates the resulting draft, applies necessary revisions, and exports the final deliverables, including the manuscript, figures, bibliography, review reports, and repair records. The detailed workflow is described in Subphases 3.1–3.6.

	\subsubsection{Subphase 3.1: Drafting Problem Formulation and Method Sections}
	Using the frozen mathematical contract from Subphase 2.1 and the frozen algorithm contract from Subphase 2.3, WritingAgent generates the formulation and method sections in LaTeX format. The resulting draft describes the wireless system model, optimization formulation, reformulation strategy, and algorithmic procedure according to the approved contracts. The controller then compiles a technical preview and verifies that the notation, equations, constraints, and algorithm descriptions remain consistent with the frozen formulation and solution route. Any inconsistencies are routed to RepairAgent for targeted revision.
	
	\begin{table*}[!t]
		\centering
		\caption{Controller Repair Routing in WARA.}
		\label{tab:repair_routing}
		\scriptsize
		\setlength{\tabcolsep}{3.5pt}
		\renewcommand{\arraystretch}{1.12}
		\begin{tabular}{p{0.14\textwidth} p{0.25\textwidth} p{0.18\textwidth} p{0.35\textwidth}}
			\hline
			\textbf{Subphase} & \textbf{Failure type} & \textbf{Artifact owner} & \textbf{Controller action} \\
			\hline
			Subphases 1.3--1.4
			& Selected wireless problem is weak or insufficiently supported
			& ScoutAgent / LiteratureAgent
			& Revise the candidate problem, expand literature search, rebuild the Phase-1 handoff, or stop with a direction-repair report. \\
			
			Subphase 2.1
			& System model or optimization formulation is inconsistent
			& FormulationAgent
			& Return the gate report for formulation repair, rerun the formulation gate, and freeze the mathematical contract only after passage. \\
			
			Subphases 2.2--2.3
			& Convexity analysis, reformulation, solver route, or algorithm description is invalid
			& TheoryAgent
			& Repair the tractability analysis or algorithm contract and rerun the theory gate before experiment code generation. \\
			
			Subphase 2.4
			& Experiment design or generated Python code fails quick validation
			& ExperimentAgent
			& Revise the experiment design or repair the generated code using harness logs and frozen contracts, then rerun the fixed validation harness. \\
			
			Subphase 2.5
			& Numerical results or figures are not paper-ready
			& ValidationAgent / ExperimentAgent
			& Request scout, medium, or paper-level reruns; route back to experiment-design or code repair; or block promotion to paper figures. \\
			
			Subphases 3.1--3.3
			& Technical writing or numerical interpretation is unsupported
			& WritingAgent / AnalysisAgent
			& Revise the \LaTeX{} text or claim--figure record using the gate report and frozen result package. \\
			
			Subphase 3.4
			& References, citation placement, or bibliography checks fail
			& LiteratureAgent / WritingAgent
			& Repair the reference bank, citation commands, citation--claim mapping, or full manuscript preview. \\
			
			Subphases 3.5--3.6
			& Final review finds a cross-artifact inconsistency
			& ReviewAgent / routed owner
			& Apply local manuscript repair or route the issue back to the responsible upstream phase before final export. \\
			\hline
		\end{tabular}
	\end{table*}

	\subsubsection{Subphase 3.2: Drafting Numerical Results and Supported Claims}
	Using the verified result package produced in Subphase 2.5, AnalysisAgent generates the numerical-results section. The generated text interprets the approved figures and experiment records and links each reported observation to its corresponding evidence source. In parallel, this subphase constructs a claim–figure traceability record that explicitly maps numerical claims to supporting figures. The controller verifies that all interpretations are evidence-grounded and remain within the scope of the validated results. Any unsupported claims are returned for revision.

	\subsubsection{Subphase 3.3: Assembly of Technical Sections}
	Using the formulation, method, and numerical-results sections, together with the supported-claim record, WritingAgent assembles the core technical manuscript. This includes generating the abstract, keywords, and conclusion while preserving consistency with the validated technical content. The controller checks completeness and verifies that the high-level narrative remains aligned with the approved model, algorithm, numerical evidence, and supported claims.

	\subsubsection{Subphase 3.4: onstruction of the Full Manuscript Draft}
	Using the assembled technical sections and the verified reference repository, LiteratureAgent first prepares validated BibTeX entries, prioritizing IEEE and peer-reviewed sources. WritingAgent then generates the introduction, positions the work relative to prior literature, and inserts citations using the verified reference keys. The controller compiles a full manuscript preview and validates citation keys, bibliography entries, venue formatting, citation placement, and citation–claim consistency. Any violations are routed to RepairAgent for correction.

	\subsubsection{Subphase 3.5: Full Manuscript Review}
	Using the complete manuscript draft together with the frozen formulation, algorithm, result, and reference artifacts, ReviewAgent performs a paper-level review. The review evaluates consistency among equations, algorithms, figures, citations, abbreviations, and technical claims. The output is a structured review report containing blocking issues, major issues, minor issues, and ownership assignments for each issue. Based on the report, the controller determines whether the problem can be resolved within Phase 3 or must be routed back to an earlier technical phase.
	
	\subsubsection{Subphase 3.6: Revision and Export of the Final Research Package}
	Guided by the review report, RepairAgent resolves manuscript issues and WritingAgent regenerates the updated manuscript. The controller then executes the final quality gate, which validates compilation, citation integrity, abbreviation consistency, figure quality, and claim support. If the manuscript passes the final gate, WARA exports the final research package, including the manuscript source, compiled PDF preview, bibliography, figures, review reports, repair history, and final validation records. Otherwise, the issue is routed either to local manuscript repair or to the responsible upstream phase, depending on its origin.

\subsection{Gate–Freezing–Repair Mechanism}
\label{subsec:gate_freezing_repair}

The gate–freezing–repair mechanism is a core component of WARA and serves as the control layer that connects all phases into a coherent research workflow. By validating artifacts before reuse, freezing approved decisions, and routing failures to responsible agents, the mechanism maintains consistency across problem formulation, algorithm design, experimentation, and manuscript generation.
\subsubsection{Gate}
A gate serves as the entry condition for downstream artifact reuse. It separates artifact generation from artifact validation: agents generate candidate artifacts, while gates evaluate whether those artifacts satisfy the active contracts and subphase requirements. Each gate produces a machine-readable report containing the artifact, pass/fail status, detected issues, and repair instructions. Passed artifacts are committed to the workspace, whereas failed artifacts are blocked from downstream use and routed to repair.

\subsubsection{Freezing} Freezing is applied to gate-approved artifacts that contain decisions required by later stages. WARA freezes four key artifacts: the problem handoff, the mathematical contract, the algorithm contract, and the verified result package. These artifacts respectively define the selected research problem, optimization formulation, solution route, and numerical evidence used throughout the remainder of the workflow. Subsequent agents may consume these artifacts but cannot modify them without triggering a repair-and-revalidation process.

\subsubsection{Repair} Repair is localized to the artifact that failed validation. Formulation errors are routed to FormulationAgent, algorithmic errors to TheoryAgent, experimental or implementation errors to ExperimentAgent, evidence-related issues to ValidationAgent, and manuscript issues to the appropriate writing or review agent. Each repair cycle records the failed gate, artifact owner, repair instructions, revised artifact, and validation outcome. If the repaired artifact passes validation, the updated version is committed and execution continues. Otherwise, WARA either performs additional repair attempts within the allocated budget or terminates the run with a structured repair report. Table \ref{tab:repair_routing} summarizes the corresponding repair-routing policy.

	\section{Case Study}
	\label{sec:case_studies}
	
	This section presents a representative WARA run from an initial wireless topic to a complete research package. The concrete run budget, including LLM output-token caps, repair limits, literature-retrieval limits, and experiment-sweep settings, is summarized below.
	
	\begin{warabox}{Case-Study Run Budget}
		\textbf{Backbone LLM:} OpenAI GPT-5.5.
		
		\smallskip
		\textbf{LLM output-token caps:} 24000 tokens for Phase 1 calls and Subphase 2.1 formulation; 18000 tokens for Subphase 2.2 tractability analysis; 10000 tokens for Subphase 2.3 algorithm design; 24000 tokens for Subphase 2.4 solver-package and validation-plan generation; 12000 tokens for Subphase 2.4 code-file and repair calls; 9000--12000 tokens for Phase 3 section-writing, numerical-writing, citation, and assembly calls; and 24000 tokens for the final ReviewAgent call.
		
		\smallskip
		\textbf{Repair limits:} one JSON-serialization repair round in Phase 1; two contract-repair rounds for Subphase 2.1--2.3; three experiment-code repair rounds and three experiment-design repair rounds in Subphase 2.4; three evidence-expansion rounds in Subphase 2.5; two technical-writing repair rounds in Subphase 3.1; and three final review-repair rounds.
		
		\smallskip
		\textbf{Literature budget:} up to eight Phase-1 literature queries with up to eight retrieved records per query, plus supplemental retrieval with up to ten records per query when needed.
		
		\smallskip
		\textbf{Experiment budget:} two promoted evidence sweeps are used: a fronthaul-capacity sweep with 12 x-axis points and 50 Monte Carlo seeds per point, and a user-load sweep with 9 x-axis points and 50 Monte Carlo seeds per point. With two compared methods, this gives 1050 paired simulation settings and 2100 method evaluations.
	\end{warabox}
	
	We next provide the main artifacts produced in the three phases of this run in the following subsections.

	\begin{figure*}[!t]
		\centering
		\fbox{%
			\begin{minipage}{1\textwidth}
				\centering
				\vspace{2pt}
				
				\includegraphics[width=0.235\textwidth]{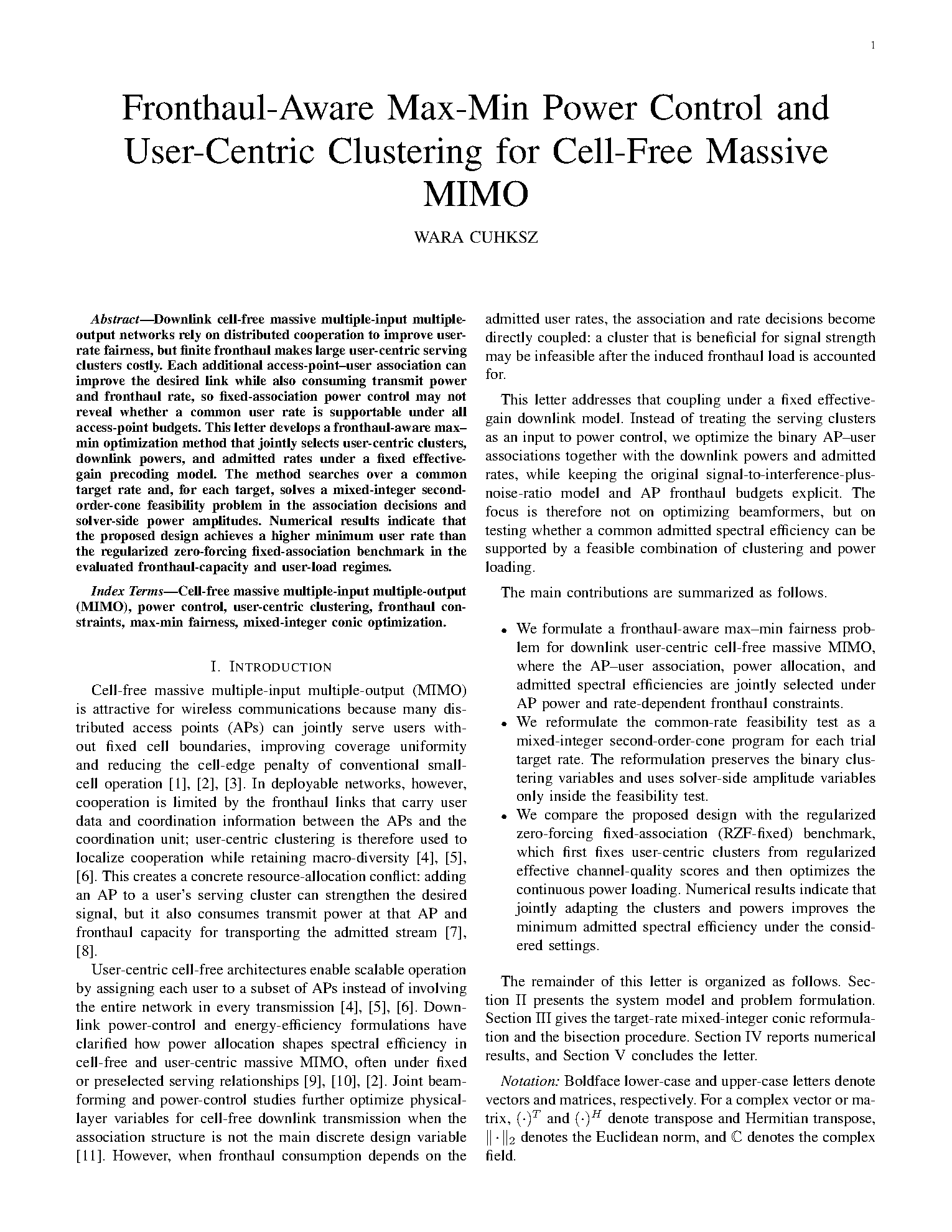}
				\hfill
				\includegraphics[width=0.235\textwidth]{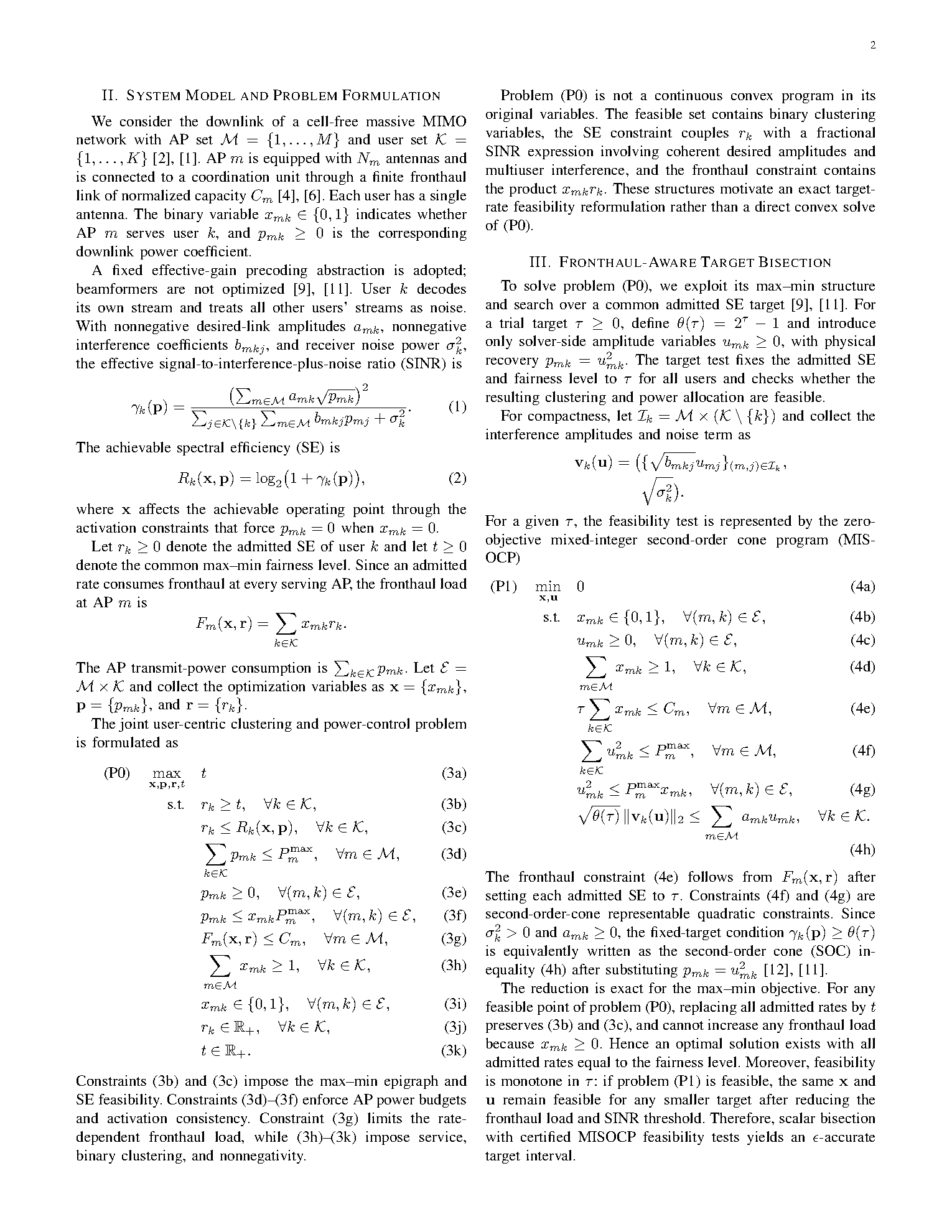}
				\hfill
				\includegraphics[width=0.235\textwidth]{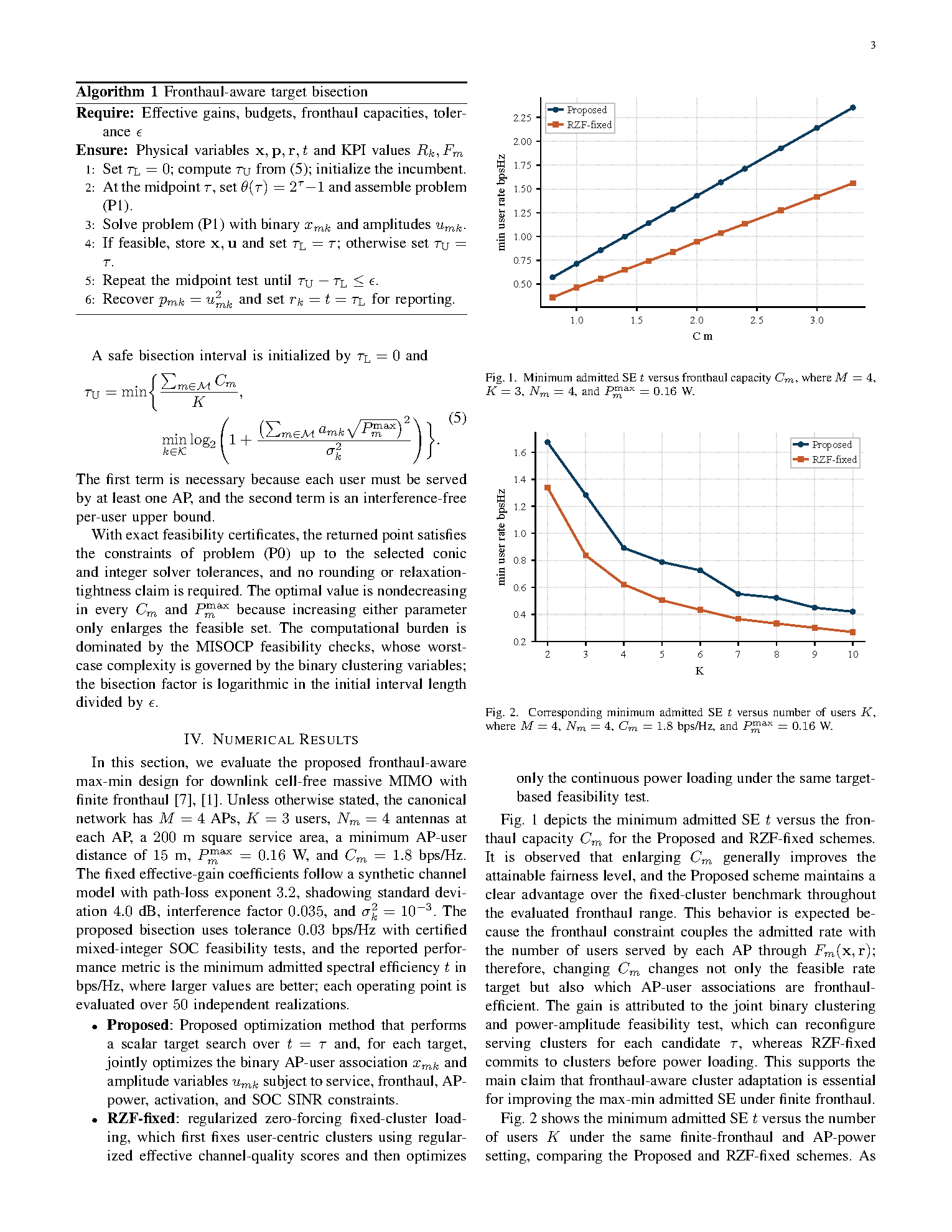}
				\hfill
				\includegraphics[width=0.235\textwidth]{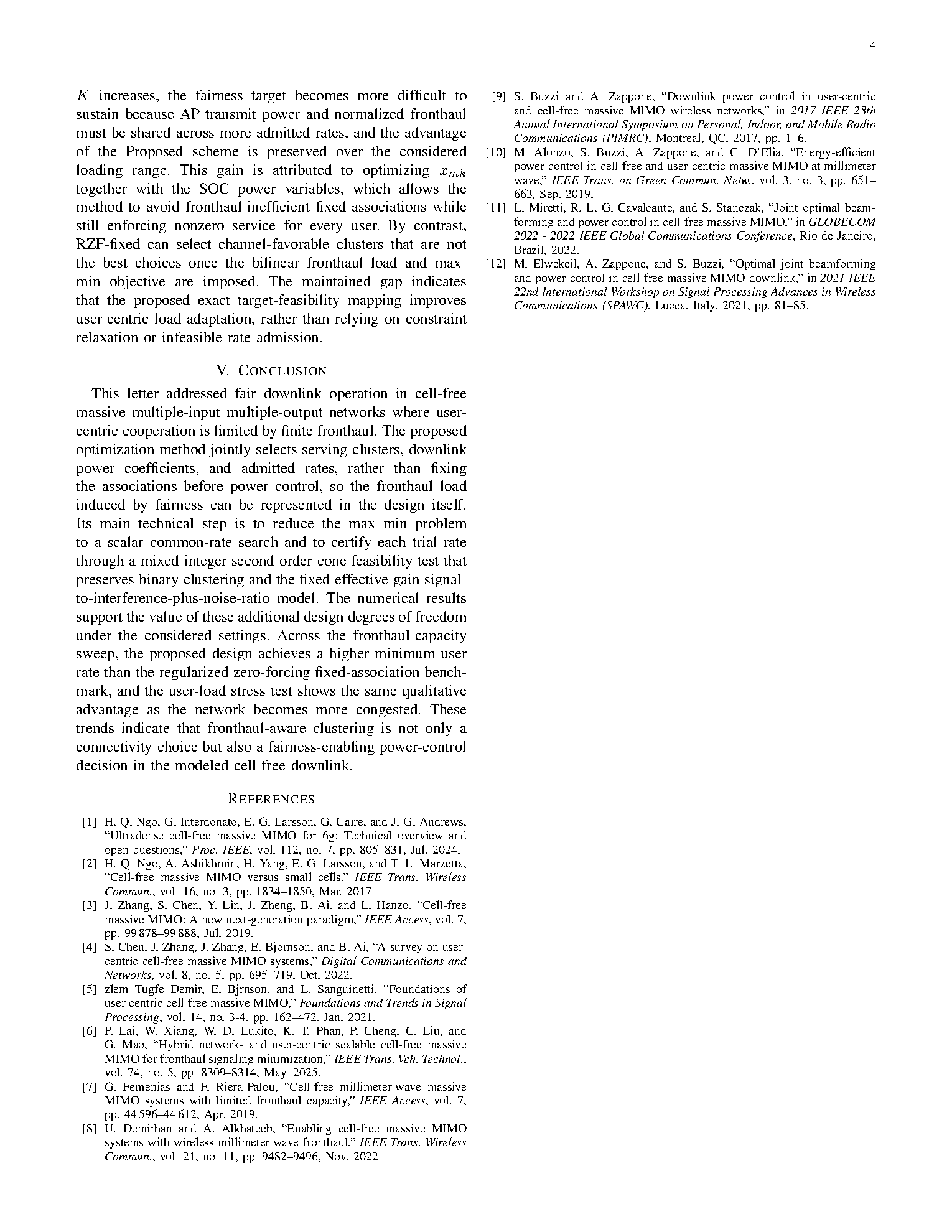}
				
			\end{minipage}
		}
		\caption{Four-page manuscript snapshot generated by WARA for the cell-free massive MIMO case.}
		\label{fig:case_paper_snapshot}
	\end{figure*}
	
	\subsection{Research Gap Identification and Problem Proposal}
	\label{subsec:case_phase1}

	The input topic is ``cell-free massive MIMO''. In Phase 1, WARA constructs and reviews candidate research directions. In this run, the selected direction is \emph{Fronthaul-Aware Max-Min Power Control and User-Centric Clustering for Cell-Free Massive MIMO}. The selected direction focuses on rate-dependent fronthaul coupling: a serving access point (AP) can improve a user's received signal, while it must also carry that user's admitted stream over a finite fronthaul link. The main Phase-1 artifact produced is summarized below, showing the selected problem direction and the information passed to the technical construction phase.

	\begin{warabox}{Phase 1 Artifact: Research Direction Contract}
		\textbf{Selected direction:} Fronthaul-aware max-min power control and user-centric clustering for downlink cell-free massive MIMO.
		
		\smallskip
		\textbf{Research object:} A downlink cell-free massive MIMO network with distributed APs, single-antenna users, finite AP transmit-power budgets, and finite per-AP fronthaul capacities.
		
		\smallskip
		\textbf{Core technical tension:} Activating an AP--user serving link can increase useful signal strength, while also consuming AP transmit power and fronthaul capacity.
		
		\smallskip
		\textbf{Selected mechanism:} Jointly optimize AP--user association and power loading so that fronthaul is allocated to serving links with high fairness value.
		
		\smallskip
		\textbf{Grounding output:} Phase 1 builds a topic-focused evidence pack covering cell-free massive MIMO foundations, user-centric clustering, downlink power control, scalable fronthaul, and fronthaul-aware cooperation.
		
		\smallskip
		\textbf{Handoff to Phase 2:} The handoff specifies the modeling seed, variables, expected max-min objective, fronthaul constraint form, validation intent, and technical risks.
	\end{warabox}

	\subsection{Wireless Optimization Modeling, Algorithm Design, and Experimentation}
	\label{subsec:case_phase2}
	
	In Phase 2, WARA develops the selected cell-free massive MIMO direction through the same modeling, solving, and experimentation pipeline described above. In Subphase 2.1, FormulationAgent turns the Phase-1 handoff into a downlink cell-free massive MIMO model and a max-min admitted-rate formulation with AP--user association, power loading, fronthaul-capacity constraints, and SINR-dependent rate constraints. In Subphases 2.2 and 2.3, TheoryAgent identifies the problem as mixed-integer and nonconvex and selects a target-rate bisection route, where each trial rate is checked through a mixed-integer conic feasibility test. In Subphase 2.4, ExperimentAgent generates the Python experiment code and runs quick validation under the fixed harness. In Subphase 2.5, WARA expands the validated experiment package and promotes two paper-ready figures: a fronthaul-capacity sweep and a user-load sweep. The main Phase-2 outputs are summarized below.
	\begin{warabox}{Phase 2 Artifact: Model, Reformulation, Algorithm, and Figures}
		\textbf{Original problem:} WARA formulates the selected problem as a fronthaul-aware max-min admitted spectral-efficiency problem for downlink cell-free massive MIMO. The binary variable $x_{mk}$ indicates whether AP $m$ serves user $k$, $p_{mk}$ is the corresponding downlink power coefficient, $r_k$ is the admitted spectral efficiency of user $k$, and $t$ is the common fairness level. The effective spectral efficiency is
		\[
		R_k(\mathbf x,\mathbf p)=\log_2\bigl(1+\gamma_k(\mathbf p)\bigr),
		\]
		where
		\[
		\gamma_k(\mathbf p)=
		\frac{\left(\sum_{m\in\mathcal M}a_{mk}\sqrt{p_{mk}}\right)^2}
		{\sum_{j\in\mathcal K\setminus\{k\}}\sum_{m\in\mathcal M}b_{mkj}p_{mj}+\sigma_k^2}.
		\]
		The fronthaul load at AP $m$ is
		\[
		F_m(\mathbf x,\mathbf r)=\sum_{k\in\mathcal K}x_{mk}r_k .
		\]
		The original problem is
		\[
		\begin{aligned}
			\text{(P0)}\quad
			\max_{\mathbf x,\mathbf p,\mathbf r,t}\quad
			& t\\
			\text{s.t.}\quad
			& r_k\ge t, && \forall k\in\mathcal K,\\
			& r_k\le R_k(\mathbf x,\mathbf p), && \forall k\in\mathcal K,\\
			& \sum_{k\in\mathcal K}p_{mk}\le P_m^{\max}, && \forall m\in\mathcal M,\\
			& p_{mk}\le x_{mk}P_m^{\max}, && \forall (m,k)\in\mathcal E,\\
			& F_m(\mathbf x,\mathbf r)\le C_m, && \forall m\in\mathcal M,\\
			& \sum\nolimits_{m\in\mathcal M}x_{mk}\ge1, && \forall k\in\mathcal K,\\
			& x_{mk}\in\{0,1\},\quad p_{mk}\ge0, && \forall (m,k)\in\mathcal E,\\
			& r_k\in\mathbb R_+, && \forall k\in\mathcal K,\quad t\in\mathbb R_+ .
		\end{aligned}
		\]
		
		\smallskip
		\textbf{Reformulated problem:} WARA exploits the max-min structure by checking a trial common target $\tau$. For each $\tau$, it sets $r_k=t=\tau$, defines $\theta(\tau)=2^\tau-1$, and introduces solver-side amplitudes $u_{mk}\ge0$ with physical recovery $p_{mk}=u_{mk}^{2}$. The target test is written as a mixed-integer second-order-cone feasibility problem over $(\mathbf x,\mathbf u)$, with binary association, service, fronthaul, AP-power, activation, and signal-to-interference-plus-noise-ratio constraints.
		
		\smallskip
		\textbf{Solution route:} Scalar bisection over $\tau$ with certified mixed-integer second-order-cone feasibility tests.
		
		\smallskip
		\textbf{Compared methods:} Proposed joint association and power-loading method versus the regularized-zero-forcing fixed-association (RZF-fixed) benchmark.
		
		\smallskip
		\textbf{Main metric:} Minimum admitted spectral efficiency $t$.
		
		\smallskip
		\textbf{Promoted figures:} Fronthaul-capacity sweep and user-load sweep.
		
		\medskip
		\centering
		\begin{minipage}{0.48\linewidth}
			\centering
			\includegraphics[width=\linewidth]{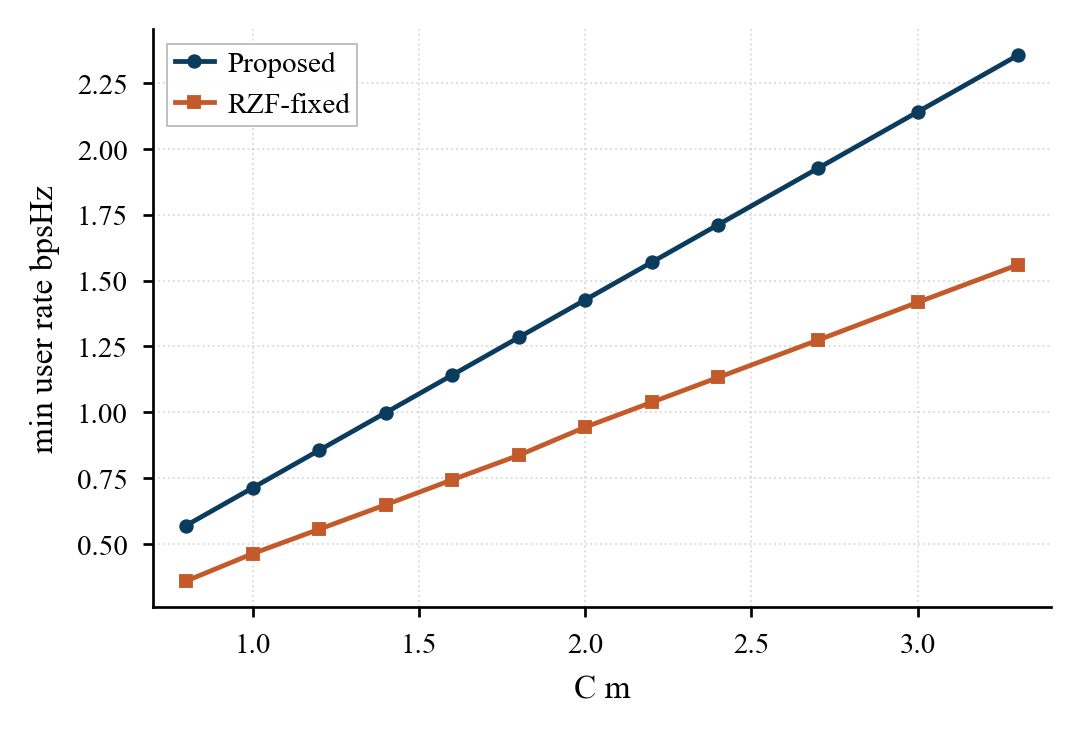}\\[-0.5mm]
			{\scriptsize Fig. (a) Fronthaul-capacity sweep.}
		\end{minipage}
		\hfill
		\begin{minipage}{0.48\linewidth}
			\centering
			\includegraphics[width=\linewidth]{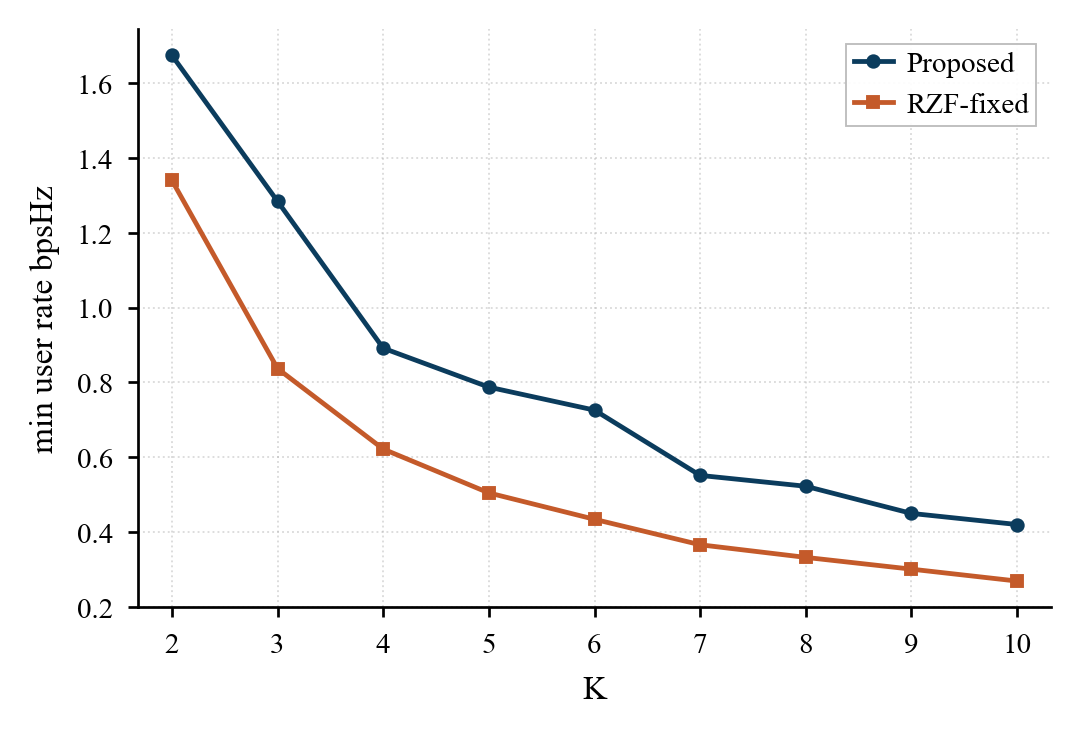}\\[-0.5mm]
			{\scriptsize Fig. (b) User-load sweep.}
		\end{minipage}
	\end{warabox}

	Fig.~(a) shows the minimum admitted spectral efficiency $t$ versus the fronthaul capacity $C_m$. The proposed method achieves a higher $t$ than the RZF-fixed benchmark because it jointly optimizes AP--user association and power loading under the fronthaul constraint. Fig.~(b) shows the minimum admitted spectral efficiency $t$ versus the number of users $K$ under the same finite-fronthaul and AP-power setting. As the user load increases, the proposed method maintains a higher fairness level than RZF-fixed by adapting the serving links and power amplitudes for each target rate, whereas RZF-fixed keeps the user-centric clusters fixed before power loading.

	\subsection{Phase 3: Research Deliverable Generation}
	\label{subsec:case_phase3}
	
	Phase 3 turns the validated Phase-2 outputs into the final paper package. In this case, WARA uses the selected problem direction, mathematical formulation, solution route, promoted figures, and verified references to generate a four-page manuscript. The final package contains 13 references, two paper-ready figures, a post-revision quality-gate report, a contract-scope report, and an abbreviation check. The generated manuscript is shown in Fig.~\ref{fig:case_paper_snapshot}.

	\begin{table*}[!t]
		\centering
		\caption{Evaluation Criteria for Manuscript-Level Research Validity and Optimization Research Maturity.}
		\label{tab:evaluation_rubric}
		\setlength{\tabcolsep}{4pt}
		\renewcommand{\arraystretch}{1.12}
		\begin{tabular}{p{0.47\textwidth} c p{0.39\textwidth}}
			\hline
			\multicolumn{3}{c}{\textbf{Manuscript-level research-validity scoring}} \\
			\hline
			\textbf{Dimension} & \textbf{Score} & \textbf{Evaluation focus} \\
			\hline
			Problem definition and scope & 10 & Clarity, boundary, and appropriateness of the research problem. \\
			Novelty and positioning & 10 & Claimed contribution and relation to prior work. \\
			System model and technical correctness & 15 & Correctness of the model, assumptions, notation, and wireless reasoning. \\
			Method validity and formulation alignment & 15 & Consistency between formulation, method, and claimed solution route. \\
			Evidence validity and experiment design & 20 & Execution provenance, baselines, scenarios, sweeps, and empirical support. \\
			Result interpretation and claim support & 15 & Whether numerical trends support the stated claims. \\
			Scientific writing and presentation & 10 & Organization, readability, figure/table quality, and technical clarity. \\
			Reference grounding & 5 & Citation relevance and grounding of prior-work claims. \\
			\hline
			\textbf{Total} & \textbf{100} & -- \\
			\hline
			
			\multicolumn{3}{c}{\textbf{Optimization research-maturity scoring}} \\
			\hline
			\textbf{Dimension} & \textbf{Scale} & \textbf{Evaluation focus} \\
			\hline
			Optimization framing & 0--100 & Wireless bottleneck, controllable resources, objective, and contribution mechanism. \\
			Formulation quality & 0--100 & Variables, assumptions, objective, and constraints of the optimization problem. \\
			Solution-method validity & 0--100 & Reformulation, algorithmic route, approximation scope, and method--formulation alignment. \\
			Benchmark design & 0--100 & Baselines, scenarios, sweeps, ablations, and fairness of comparison. \\
			Evidence strength & 0--100 & Stability, interpretability, and sufficiency of numerical evidence. \\
			Scholarly positioning & 0--100 & Relation to prior wireless optimization literature. \\
			\hline
			\textbf{Overall} & \textbf{Mean} & Arithmetic mean of the six normalized dimension scores. \\
			\hline
		\end{tabular}
	\end{table*}
	\section{Comparative Evaluation Results}
	\label{sec:results}
	
	In this section, we evaluate the research quality of WARA-generated manuscripts using a structured LLM-based \emph{scoring agent} as evaluator. The evaluation has two purposes. First, it measures the overall validity of the final manuscript, including problem definition, novelty, technical correctness, method--formulation alignment, evidence support, claim consistency, writing quality, and reference grounding. Second, it zooms in on the optimization part of the manuscript, because wireless optimization papers require not only readable exposition but also a coherent chain from wireless bottleneck, mathematical formulation, solution method, benchmark design, and numerical evidence. We therefore report both a manuscript-level research-validity comparison and an optimization research-maturity comparison.

	\subsection{ScoringAgent and Evaluation Criteria}
	\label{subsec:scoring_agent}
	
	We design a structured LLM-based {ScoringAgent} to evaluate the research quality of manuscripts. For each evaluated manuscript, ScoringAgent receives a manuscript PDF and calls an evaluator LLM with a fixed reviewer-style prompt. The prompt instructs the LLM to act as a strict wireless-communications reviewer, apply predefined scoring criteria, justify each score using manuscript-visible evidence, and return a machine-readable JSON report. The report contains dimension-level scores, brief justifications, overall scores, identified strengths and weaknesses, and a confidence value. The same scoring prompts, scoring criteria, output schema, parsing rules, and score-range checks are used for all evaluated manuscripts.
	
	The evaluation uses two complementary scoring criteria, as summarized in Table~\ref{tab:evaluation_rubric}. The first criterion measures manuscript-level research validity. It evaluates whether the manuscript defines a clear research problem, positions its contribution, presents a technically correct system model, aligns the method with the formulation, provides valid evidence, supports its claims, communicates clearly, and grounds its statements in references. This criterion is the main aggregate quality score and uses an additive 100-point scale. The second criterion measures optimization research maturity. It is used as a diagnostic profile for the optimization part of the study, focusing on optimization framing, formulation quality, solution-method validity, benchmark design, evidence strength, and scholarly positioning. Each optimization-maturity dimension is scored on a normalized 0--100 scale, and the overall optimization-maturity score is computed as the arithmetic mean of the six dimension scores.
	
	For the $i$-th manuscript $\mathcal P_i$, ScoringAgent first assigns manuscript-level dimension scores
	\begin{equation}
		\mathbf s_i^{\mathrm R}
		=
		[s_{i,1}^{\mathrm R},s_{i,2}^{\mathrm R},\ldots,s_{i,8}^{\mathrm R}],
	\end{equation}
	according to the eight research-validity dimensions in Table~\ref{tab:evaluation_rubric}. The manuscript-level research-validity score is computed as
	\begin{equation}
		S_i^{\mathrm R}
		=
		\sum_{d=1}^{8}s_{i,d}^{\mathrm R}.
	\end{equation}
	
	For optimization-focused evaluation, ScoringAgent also assigns a six-dimensional maturity profile
	\begin{equation}
		\mathbf u_i
		=
		[u_{i,1},u_{i,2},\ldots,u_{i,6}],
		\qquad 0\le u_{i,d}\le 100 ,
	\end{equation}
	where each entry is a normalized score for one optimization-maturity dimension. Consistent with Table~\ref{tab:evaluation_rubric}, the overall optimization research-maturity score is
	\begin{equation}
		S_i^{\mathrm O}
		=
		\frac{1}{6}\sum_{d=1}^{6}u_{i,d}.
	\end{equation}
	
	For a benchmark set $\mathcal B$ with $N$ manuscripts, we report the average manuscript-level score
	\begin{equation}
		\bar S^{\mathrm R}(\mathcal B)
		=
		\frac{1}{N}\sum_{\mathcal P_i\in\mathcal B}S_i^{\mathrm R}.
	\end{equation}
	We also report the average optimization research-maturity score
	\begin{equation}
		\bar S^{\mathrm O}(\mathcal B)
		=
		\frac{1}{N}\sum_{\mathcal P_i\in\mathcal B}S_i^{\mathrm O}.
	\end{equation}

	\subsection{Benchmark Sets and Comparative Results}
	\label{subsec:benchmark_results}
	
	We compare three benchmark sets. The WARA and one-shot LLM sets are topic-paired, while the accepted IEEE Wireless Communications Letters (WCL) set is used as a peer-reviewed reference profile. Specifically, the benchmark sets are:
	
	\begin{itemize}
		\item \textbf{WARA-generated manuscripts:} ten manuscripts produced by the proposed closed-loop WARA workflow from ten wireless optimization topics, using OpenAI GPT-5.5 as the backbone LLM.
		\item \textbf{One-shot LLM manuscripts:} ten manuscripts generated from the same ten topics using a single prompt and the same OpenAI GPT-5.5 backbone, without phase-structured  artifact control, executable validation, or repair.
		\item \textbf{Accepted WCL papers:} ten randomly selected, recently accepted optimization-related IEEE Wireless Communications Letters papers used as a peer-reviewed reference group.
	\end{itemize}
	
	To reduce evaluator--generator coupling, the scoring agent is implemented using Kimi K2.6 rather than the GPT-5.5 backbone used for the generated manuscripts.

	\begin{table*}[!t]
		\centering
		\caption{Manuscript-Level Validity Scores.}
		\label{tab:research_validity_scores}
		\setlength{\tabcolsep}{3.2pt}
		\renewcommand{\arraystretch}{1.12}
		\begin{tabular}{l c c c c c c c c c c}
			\hline
			\textbf{Benchmark set} & \textbf{$N$} & \textbf{Overall} & \textbf{Scope} & \textbf{Novelty} & \textbf{Model} & \textbf{Method} & \textbf{Evidence} & \textbf{Claims} & \textbf{Writing} & \textbf{Refs} \\
			& & \textbf{/100} & \textbf{/10} & \textbf{/10} & \textbf{/15} & \textbf{/15} & \textbf{/20} & \textbf{/15} & \textbf{/10} & \textbf{/5} \\
			\hline
			One-shot LLM & 10 & 37.4$\pm$2.7 & 6.0 & 5.7 & 7.9 & 6.7 & 0.0 & 3.0 & 5.0 & 3.3 \\
			WARA & 10 & 68.5$\pm$5.7 & 8.0 & 6.8 & 10.4 & 9.8 & 13.0 & 9.3 & 6.9 & 3.6 \\
			Accepted WCL & 10 & 81.4$\pm$4.1 & 8.4 & 7.6 & 12.2 & 11.7 & 17.1 & 12.3 & 8.0 & 4.2 \\
			\hline
		\end{tabular}
	\end{table*}

	\begin{figure}[!t]
		\centering
		\includegraphics[width=1\linewidth]{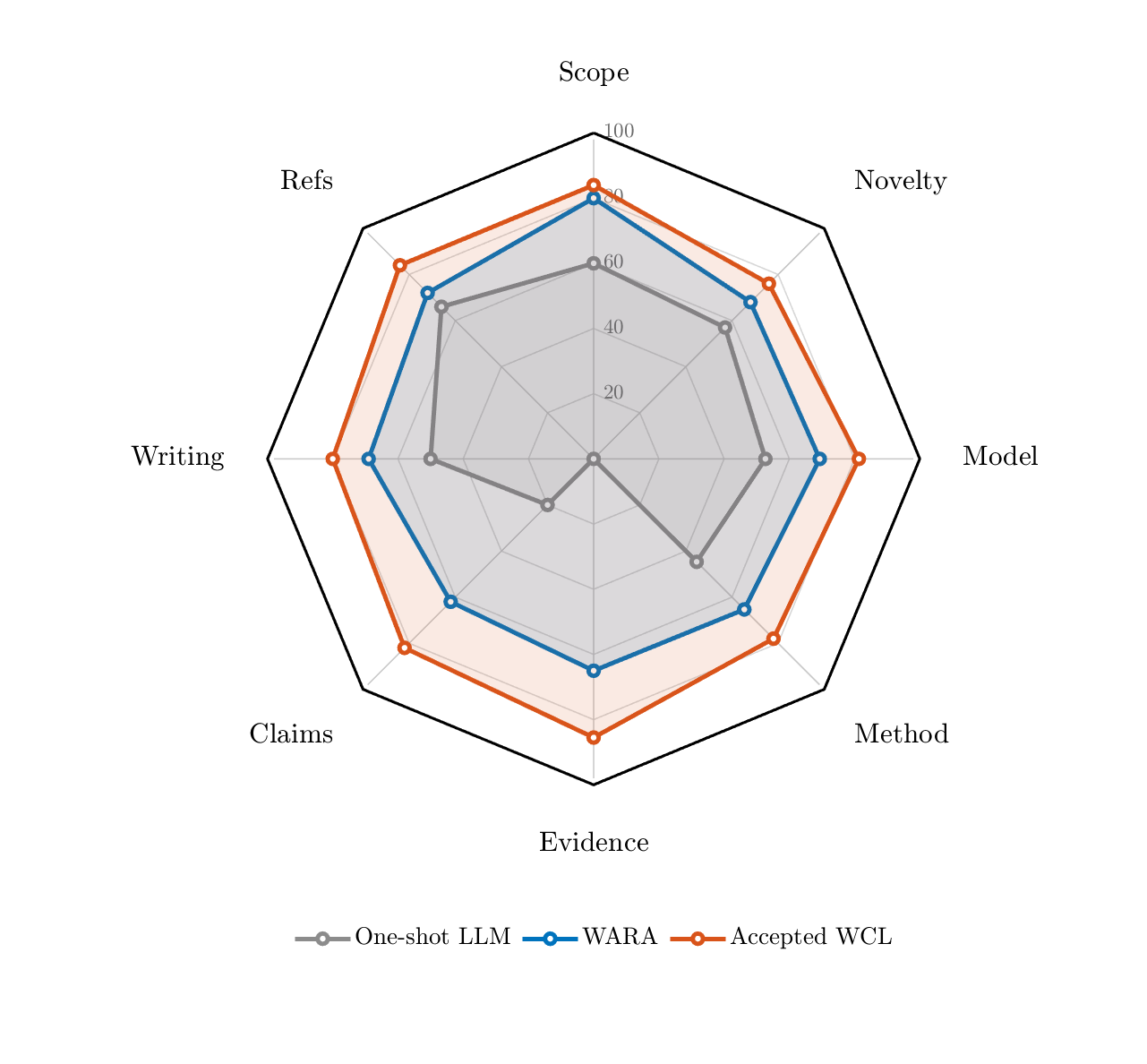}
		\caption{Manuscript-level validity profiles.}
		\label{fig:dimension_profile}
	\end{figure}
	\begin{figure}[!t]
		\centering
		\includegraphics[width=1\linewidth]{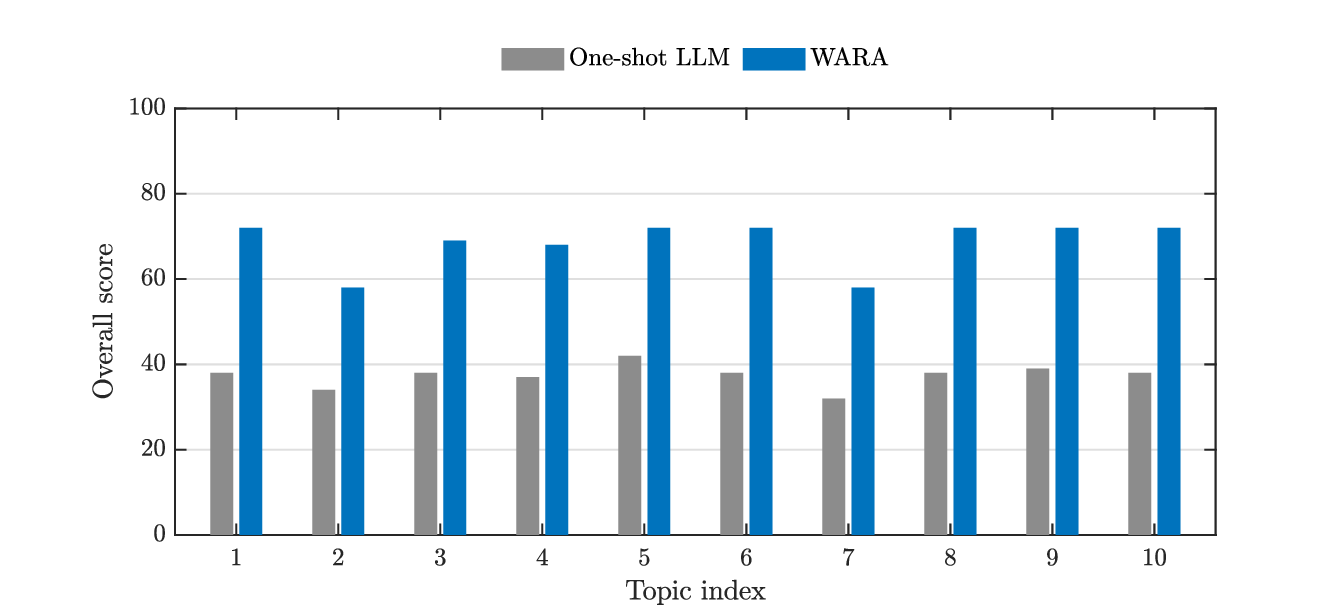}
		\caption{Topic-level paired manuscript scores for WARA and the one-shot LLM baseline.}
		\label{fig:manuscript_scores}
	\end{figure}
	Table~\ref{tab:research_validity_scores} and Fig.~\ref{fig:dimension_profile} report the manuscript-level research-validity comparison. WARA obtains an average score of 68.5, compared with 37.4 for the topic-paired one-shot LLM baseline, yielding a 31.1-point improvement under the same topic set and backbone model. The largest gap appears in evidence validity, where the one-shot baseline receives 0.0 because its numerical results are generated as text without recorded experiment execution, while WARA reaches 13.0 through executable validation and evidence packaging. WARA also improves claim support from 3.0 to 9.3, showing that its performance claims are more directly tied to validated figures and recorded experiment outputs. These two dimensions benefit most from closed-loop control because WARA explicitly separates experiment generation, execution, evidence verification, and claim writing into different gated stages. As a result, unsupported numerical descriptions are blocked before synthesis, and downstream claims must remain within the scope of the verified evidence. The gains in system-model correctness and method validity further show that staged formulation and algorithm construction help reduce mismatch between the stated optimization problem and the proposed solution. Compared with accepted WCL papers, WARA is closer in problem definition and novelty, but still weaker in evidence validity, claim support, and method maturity. This suggests that WARA improves the reliability of generated manuscripts mainly by making the research process executable, while the remaining gap comes from the depth and refinement of human-designed experiments and technical arguments.
	
	Fig.~\ref{fig:manuscript_scores} reports the topic-level paired comparison between WARA and the one-shot LLM baseline. WARA outperforms the corresponding one-shot manuscript on all ten topics, showing that the gain is not caused by a few favorable examples. The reason is that WARA does not rely on a single forward generation from topic to manuscript. Instead, each topic is repeatedly interpreted, checked, and revised through multiple agent stages. In specific, the research direction is reviewed before formulation, the formulation is frozen before algorithm design, the algorithm is checked before implementation, and the evidence is validated before writing. When a gate fails, the repair mechanism sends the problem back to the responsible stage instead of letting the final manuscript absorb the inconsistency. This multi-stage reasoning and targeted repair process explains why WARA produces more stable improvements across different wireless optimization topics.

\begin{figure}[!t]
		\centering
		\includegraphics[width=1\linewidth]{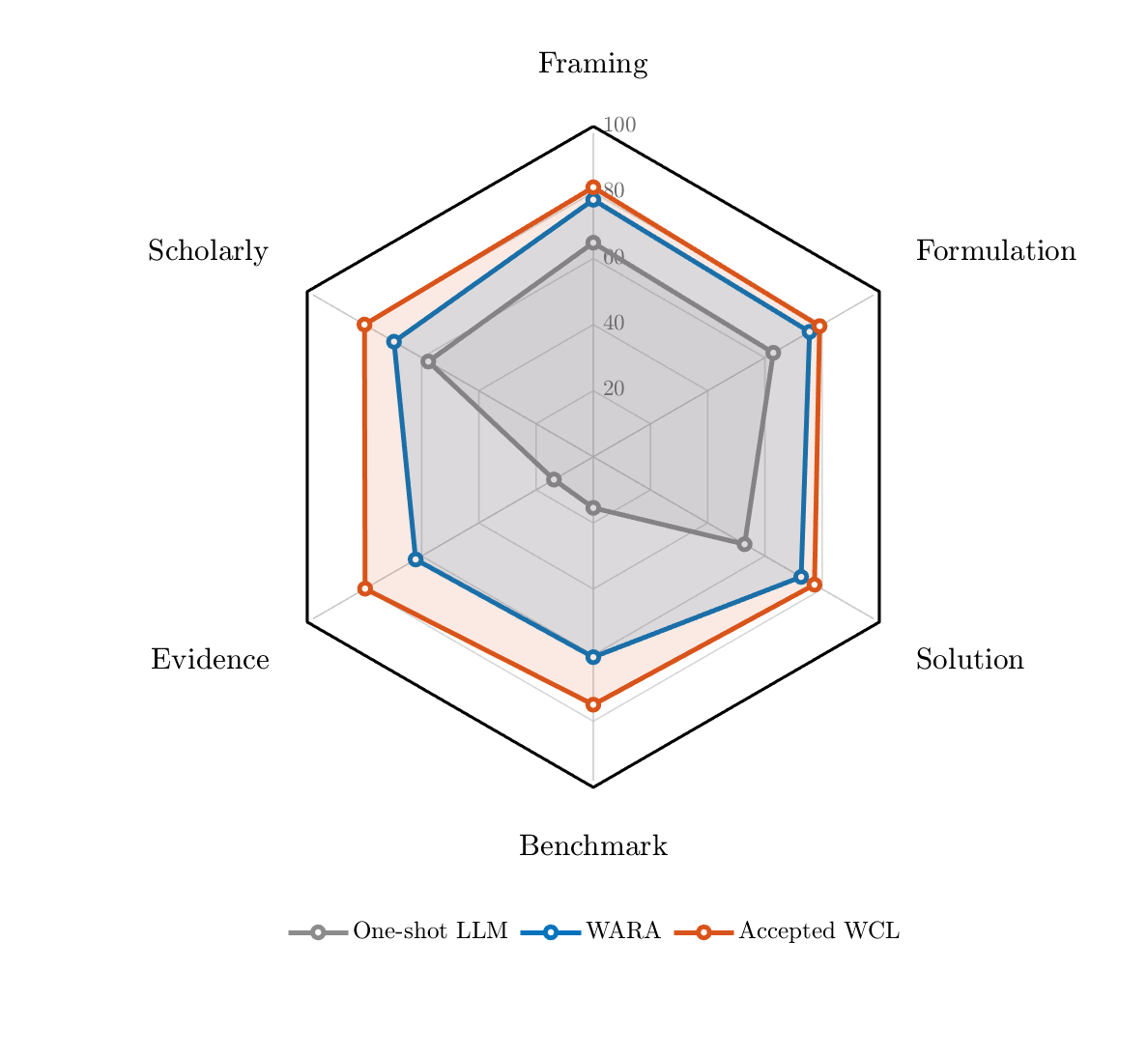}
		\caption{Optimization-maturity profiles.}
		\label{fig:optimization_maturity_profile}
	\end{figure}

	Table~\ref{tab:optimization_maturity_scores} and Fig.~\ref{fig:optimization_maturity_profile} report the optimization-focused comparison. WARA improves the average optimization research-maturity score from 44.6 to 69.7 over the one-shot baseline. The key reason is that WARA treats optimization as a structured problem-solving process. It first identifies the wireless resource-coupling mechanism, then fixes the variables, objective, constraints, and assumptions as a mathematical contract, and only then derives a solution route and experiment design from that contract. This makes the proposed method and benchmarks accountable to a specific optimization problem. In contrast, one-shot generation tends to produce the formulation, method, and evaluation in a single step, so the optimization components are less likely to be mutually checked. This explains why WARA improves not only evidence strength, but also formulation quality, solution-method validity, and benchmark design. The accepted WCL papers still obtain the highest score of 78.8, mainly because published optimization studies usually contain more mature baseline selection, sharper ablation logic, stronger scenario design, and more carefully calibrated claims. \textcolor{black}{This gap indicates that WARA can already construct a clearer optimization research structure, but still lacks the full experimental depth and contribution-level judgment of mature human-authored wireless optimization studies.}

	\begin{table*}[!t]
		\centering
		\caption{Optimization-Maturity Scores. }
		\label{tab:optimization_maturity_scores}
		\setlength{\tabcolsep}{6.0pt}
		\renewcommand{\arraystretch}{1.12}
		\begin{tabular}{l c c c c c c c c}
			\hline
			\textbf{Benchmark set} & \textbf{$N$} & \textbf{Overall} & \textbf{Framing} & \textbf{Formulation} & \textbf{Solution} & \textbf{Benchmark} & \textbf{Evidence} & \textbf{Scholarly} \\
			& & \textbf{/100} & \textbf{/100} & \textbf{/100} & \textbf{/100} & \textbf{/100} & \textbf{/100} & \textbf{/100} \\
			\hline
			One-shot LLM & 10 & 44.6 & 64.8 & 62.9 & 52.9 & 15.4 & 13.8 & 57.7 \\
			WARA & 10 & 69.7 & 77.8 & 75.6 & 72.7 & 60.6 & 62.1 & 69.7 \\
			Accepted WCL & 10 & 78.8 & 81.6 & 79.1 & 77.3 & 75.0 & 79.8 & 80.0 \\
			\hline
		\end{tabular}
	\end{table*}
	
	\section{Conclusion}
	\label{sec:conclusion}
	
	This paper presented WARA, a closed-loop multi-agent framework for automated wireless optimization research. The central idea is to treat a wireless research study as a chain of verifiable artifacts rather than as a single generated manuscript. Starting from an initial topic, WARA constructs a research direction, develops the corresponding optimization study, generates executable evidence, and synthesizes the final research deliverables. Controller-managed gates, frozen interfaces, and targeted repair are used to keep these artifacts consistent as the research package evolves.
	
	The case study shows how WARA turns a broad wireless topic into a concrete optimization study with a bounded direction, a frozen mathematical contract, an implementable solution route, and verified numerical evidence. The comparative evaluation further shows that WARA improves substantially over one-shot LLM generation. The gains are most visible in formulation--method alignment, evidence validity, claim support, and optimization research maturity. These results indicate that LLM-agent-assisted wireless research benefits from phase-structured artifact construction and executable feedback, rather than relying only on fluent end-to-end generation.
	
	The remaining gap to accepted peer-reviewed wireless papers lies mainly in experimental depth and evidence maturity. WARA can build a consistent research package and produce executable numerical results, but mature wireless optimization papers usually require stronger experimental judgment.  A promising direction is therefore to integrate human expertise into WARA as part of the artifact chain rather than as external comments. Expert feedback can be written back as revised contracts, gate criteria, benchmark requirements, simulator settings, and evidence-to-claim rules, which later agents can directly read and enforce. This would allow WARA to reuse expert decisions across runs, update its benchmark and simulator libraries, and request human input only when automatic checks cannot resolve contribution strength, baseline adequacy, or evidence sufficiency.

	\bibliographystyle{IEEEtran}
	\bibliography{ref}
	
\end{document}